\documentclass[%
  prb,
  aps,
  twocolumn,
  reprint,
  superscriptaddress,
  longbibliography,
  floatfix,
]{revtex4-1}

\usepackage{amsmath,amssymb,units,txfonts}
\usepackage{amsfonts}

\usepackage{graphicx}
\usepackage{helvet,wasysym}
\usepackage[bookmarks=true,colorlinks=true,urlcolor=blue,linkcolor=blue,citecolor=blue]{hyperref} 
\usepackage[usenames,dvipsnames]{color}
\usepackage{colortbl,cuted,bm}
\usepackage{multirow}
\usepackage{xr}
\usepackage{lineno}
\usepackage{multirow}
\usepackage{booktabs}
\usepackage{lipsum}

\newcommand{\bk}{{\bf k}}
\newcommand{\br}{{\bf r}}
\newcommand{\bq}{{\bf q}}
\newcommand{\hbq}{\hat{\bf q}}
\newcommand{\bG}{{\bf G}}

\newcommand{\Proj}{\mathcal{T}}

\newcommand{\rv}{\textbf{r}}
\newcommand{\BigO}{\mathcal{O}}

\newcommand{\Deltax}{\Delta_{\textit{x}}}

\newcommand{\rhoex}{\rho_{\textit{ex}}}
\newcommand{\LCAOTDDFTkomega}{LCAO-TDDFT-\emph{k}-$\omega$}
\newcommand{\RSTDDFTkomega}{RS-TDDFT-\emph{k}-$\omega$}
\newcommand{\PWTDDFTkomega}{PW-TDDFT-\emph{k}-$\omega$}
\newcommand{\TDDFTromega}{TDDFT-\emph{r}-$\omega$}
\newcommand{\RSTDDFTrt}{RS-TDDFT-\emph{r}-\emph{t}}
\newcommand{\GLLBsc}{GLLB-SC}
\newcommand{\Gr}{\textsc{Gr}}
\newcommand{\Pn}{\textsc{Pn}}
\newcommand{\AlO}{Al$_{\text{2}}$O$_{\text{3}}$}
\newcommand{\RTiO}{R-TiO$_{\text{2}}$}
\newcommand{\ATiO}{A-TiO$_{\text{2}}$}

\newcommand{\Ecut}{E_{\textit{cut}}}
\newcommand{\temp}{k_{\text{B}}T}
\newcommand{\EgapQP}{E_{\text{gap}}^{\text{QP}}}
\newcommand{\Ebind}{E_{\text{bind}}}

\usepackage[normalem]{ulem}
\usepackage[utf8x]{inputenc}
\usepackage{xfrac}
\DeclareMathOperator{\Imag}{Im}
\DeclareMathOperator{\Real}{Re}

\begin{document}
\title{LCAO-TDDFT-\emph{k}-$\omega$: Spectroscopy in the Optical Limit}
\author{Keenan Lyon}
\affiliation{Department of Applied Mathematics, University of Waterloo, Waterloo, Ontario, Canada}
\author{Mar\'{\i}a Rosa Preciado-Rivas}
\affiliation{School of Physical Sciences and Nanotechnology, Yachay Tech University, Urcuqu\'{\i} 100119, Ecuador}
\author{Duncan John Mowbray}
\affiliation{School of Physical Sciences and Nanotechnology, Yachay Tech University, Urcuqu\'{\i} 100119, Ecuador}
\affiliation{Department of Applied Mathematics, University of Waterloo, Waterloo, Ontario, Canada}
\email{duncan.mowbray@gmail.com}
\author{Vito Despoja}
\affiliation{Institute of Physics, Bijeni\u{c}ka 46, HR-10000 Zagreb, Croatia}

\begin{abstract} 
  Understanding, optimizing, and controlling the optical absorption process, exciton gemination, and electron-hole separation and conduction in low dimensional systems is a fundamental problem in materials science.  However, robust and efficient methods capable of modelling the optical absorbance of low dimensional macromolecular systems and providing physical insight into the processes involved have remained elusive.  We employ a highly efficient linear combination of atomic orbitals (LCAOs) representation of the Kohn--Sham (KS) orbitals within time dependent density functional theory (TDDFT) in the reciprocal space ($k$) and frequency ($\omega$) domains, as implemented within our LCAO-TDDFT-$k$-$\omega$ code, and apply the derivative discontinuity correction of the exchange functional $\Deltax$ to the KS eigenenergies. In so doing we are able to provide a semi-quantitative description of the optical absorption, conductivity, and polarizability spectra for prototypical 0D, 1D, 2D, and 3D systems within the optical limit ($\|\bq\|\to0^+$) as compared to both available measurements and from solving the Bethe--Salpeter equation with quasiparticle $G_0 W_0$ eigenvalues ($G_0 W_0$-BSE).  Specifically, we consider 0D fullerene (C$_{60}$), 1D metallic (10,0) and semiconducting (10,10) single-walled carbon nanotubes (SWCNTs), 2D graphene (\textsc{Gr}) and phosphorene (\textsc{Pn}), and 3D rutile (R-TiO$_2$) and anatase (A-TiO$_2$).  For each system, we also employ the spatially resolved electron-hole density to provide direct physical insight into the nature of their optical excitations.  These results demonstrate the reliability, applicability, efficiency, and robustness of our LCAO-TDDFT-$k$-$\omega$ code, and open the pathway to the computational design of macromolecular systems for optoelectronic, photovoltaic, and photocatalytic applications \emph{in silico}.
\end{abstract}

\maketitle


\section{Introduction}\label{Introduction}

The understanding, optimization, and control of optical absorption processes, exciton generation and recombination, and electron-hole separation and conduction is a basic challenge in nanomaterial design, with a plethora of applications in optoelectronics\cite{Optoelectronics2004,Optoelectronics2012,MolecularPlasmonicsACSPhotonics2019}, photovoltaics\cite{AngelFullerenesPhotovoltaics,PolymerFullerenePhotovoltaic,FullereneOnlyPhotovoltaic,Fullerene_apl1,berson2007elaboration,PolymerSolarCellsRev,DyeSensitizedSolarCellsRev}, and photocatalysis\cite{TiO2PhotocatalysisChemRev2014,OurJACS}.  Functionalizable and tailorable optically active low-dimensional nanomaterials, such as fullerenes (C$_{60}$)\cite{Fullerene_apl1,AngelFullerenesPhotovoltaics,PolymerFullerenePhotovoltaic,FullereneOnlyPhotovoltaic,berson2007elaboration}, single-walled carbon nanotubes (SWCNTs)\cite{Fullerene_apl1,berson2007elaboration,kymakis2002single,campidelli2008facile,bartelmess2010phthalocyanine}, graphene (\Gr), phosphorene (\Pn) \cite{phosphorene,materials-database}, and rutile (\RTiO) and anatase (\ATiO) nanoparticles \cite{TiO2PhotocatalysisChemRev2014}, have attracted particular interest as prototypical optoelectronic, photovoltaic, and photocatalytic systems.  However, robust and efficient theoretical methods that take into account the peculiarities of modelling optical absorption in low dimensional systems\cite{RadialCutoff,BenedictNanotubes,DobsonLayer,Matthes2016,Mowbray2014} are somewhat scarce.  


\emph{Ab initio} time dependent density functional theory (TDDFT) methods\cite{Maitra,Ullrich2011TDDFTbook} are the standard approaches for modelling light--matter interactions. They run the gamut from highly efficient but often qualitatively inaccurate methods based on the Kohn--Sham (KS) density of states \cite{Kohn_1965} to quantitatively accurate but computationally inefficient methods based on the hybrid exchange and correlation (xc) functional (HSE)\cite{HSE} or quasiparticle ($GW$)\cite{GW} electronic structure and the solution to the four-point Bethe--Salpeter equation (BSE)\cite{Bethe_1951}.  


While TDDFT in real space ($r$) and frequency ($\omega$) domains (\TDDFTromega{})  \cite{Casida1995, TDDFTRevCasida2009} often provides the desired balance between accuracy and efficiency, both its restriction to non-periodic (0D) calculations and $\BigO(N^5)$ scaling make it unsuitable for modelling low dimensional (1D or 2D) systems and macromolecules.  With a real space (RS) representation of the KS wave functions, one can both perform systematic convergence with respect to the grid spacing $h$ and accurately describe both the local and non-local features which are inherent in low dimensional systems.  In this case, TDDFT in real space ($r$) and time ($t$) domains (\RSTDDFTrt{})\cite{Yabana_1996, Koh_2017} tends to require unnecessarily short time steps to yield stability of the wavefunctions, much shorter than the time steps required to resolve the frequency spectra \cite{Jornet_2015}, whereas TDDFT in reciprocal space $k$ and frequency $\omega$ domains (\RSTDDFTkomega{}) \cite{response1, response2} has quite high memory costs and its accuracy strongly depends on an appropriate choice of xc functional.

A plane wave (PW) representation of the KS wavefunctions reduces the computational cost and improves stability relative to a RS representation while still allowing a systematic convergence with respect to the PW energy cutoff $\Ecut$.  However, such a representation is rather unsuitable for non-periodic or mixed boundary condition (0D, 1D, or 2D) systems. Moreover, although \PWTDDFTkomega{} calculations can heavily leverage fast Fourier transforms (FFTs) in their execution, their high memory cost can often make them unfeasible, especially for systems with large unit cells.

  The $G_0 W_0$-BSE method has proven to be one of the most quantitatively accurate methods for modelling one- and two-particle excitations. Based on a systematic perturbation theory approach around the weaker screened Coulomb interaction, $G_0 W_0$ includes the screening of the electronic band structure through the frequency-dependent self energy $\Sigma(\omega)$\cite{GW,Huser_2013}. The resulting quasiparticle electronic levels\cite{PlasmonPole1995} are thus corrected to account for their screening by the media, with unoccupied levels typically upshifted to increase the electronic band gap.  To describe the optical band gap, exciton binding, i.e., electron-hole interactions, should also be included through the solution of the four-point Bethe-Salpeter equation\cite{Bethe_1951} when calculating the macroscopic dielectric function. Computationally, a $G_0W_0$ calculation will scale cubically with the number of plane waves\cite{GWPAW}, whereas constructing and applying the Bethe-Salpeter Hamiltonian will scale with the $k$-point sampling of the Brillouin Zone (BZ), resulting in restrictive time and memory costs\cite{BSEscaling}. Although neither electronic screening nor excitonic binding are included in standard TDDFT calculations, these effects tend to compensate for each other in the calculated $G_0 W_0$-BSE optical absorbance spectra, often leading to a fortuitous error cancellation with TDDFT. This suggests we should use $G_0 W_0$-BSE calculations to benchmark less computationally intensive TDDFT methods where both quasiparticle screening and strong excitonic effects are neglected.

In this work, we use linear combinations of atomic orbitals (LCAOs)\cite{GPAWLCAO} to represent the KS wavefunctions, and perform TDDFT calculations in reciprocal space ($k$) and frequency ($\omega$) domains (\LCAOTDDFTkomega).  In so doing, we provide an efficient and more stable representation of the KS orbitals while obtaining a similar accuracy to \PWTDDFTkomega{}.  However, since we cannot efficiently apply an FFT within our LCAO representation of the KS orbitals, we are restricted in our \LCAOTDDFTkomega{} calculations to the head of the dielectric function, and the optical limit $\|\bq\| \to 0^+$.

In order for \LCAOTDDFTkomega{} to obtain accurate optical absorption spectra, we require an accurate description of the system's electronic structure in line with experimentally measured band gaps \cite{Emery_2016}.  To do so, we employ the derivative discontinuity correction to the exchange (x) part of the \GLLBsc{}\cite{GLLB} functional to perform an \emph{ab initio} upshift of the unoccupied KS eigenenergies.  This approach already successfully described the optical absorbance of both 0D chlorophyll monomers\cite{Chlorophyll} and 1D SWCNTs\cite{PreciadoSWCNTs}.

LCAO representations are inherently reliant on the choice of basis set, and unlike PW or RS calculations, systematic convergence cannot be obtained by simply decreasing the grid spacing $h$ or increasing the plane-wave cutoff energy $E_{\textit{cut}}$. The basis set choice is even more important for TDDFT relative to ground state DFT, as the basis set must describe both occupied and unoccupied states equally well \cite{Kuisma_2015}. The importance of benchmarking for any LCAO basis set method is thus quite evident.

Herein we compare and contrast for a set of prototypical low-dimensional optically active nanomaterials the response functions and spectra obtained from different LCAO basis sets, \PWTDDFTkomega{} and $G_0 W_0$-BSE calculations, and experimental measurements.  Both $p$-valence and completeness-optimized basis sets have been put forward as methods for improving the density of states of unoccupied states and resolving the absolute convergence issue \cite{Kuisma_2015, LCAOBasisSets}.  However, since these methods have not been tested on a sufficiently wide range of materials, we shall restrict consideration to the default LCAO basis sets and directly compare to PW representations of the KS wavefunctions herein\cite{GPAWLCAO}.

As a consequence of having derived quantities independent of spatial dispersion, our \LCAOTDDFTkomega{} calculations have direct access to the transitions involved at a particular energy. This allows the description of the spatially-resolved excited electron and hole densities, providing physical insight into the excitations at different energies. In addition, the oscillator strengths themselves, related to the dipole transition matrix \cite{Adler}, provide insight into the reciprocal-space distribution of the transitions within the BZ of the material at a given energy.

This paper is organized as follows.  In Sec.~\ref{Methodology} we provide theoretical background to \LCAOTDDFTkomega, low-dimensional response functions, and real-space electron and hole densities, followed by a complete description of the relevant computational parameters employed herein. To highlight the range of applicability of the \LCAOTDDFTkomega{} method, in Sec.~\ref{results} we apply it to C$_{60}$  (0D), metallic (10,10) and semiconducting (10,0) SWCNTs (1D), \Gr{} and \Pn{} monolayers (2D), and bulk \RTiO{} and \ATiO{} (3D), performing comparisons with \PWTDDFTkomega{}, $G_0W_0$-BSE, and experimental measurements, and showcasing its spatially-resolved electron-hole density difference. Concluding remarks are given in Sec.~\ref{Sect:Conclusions} followed by an Appendix with further details of the derivative discontinuity correction  $\Deltax$, the \LCAOTDDFTkomega{} method's treatment of the BZ, the $k$-point convergence of \Gr{} and \Pn{}'s in-plane conductivity, and the derived model and measurements of \Pn's reflection spectra.  Atomic units ($\hslash = e = m_e = a_0 = 1$) have been employed throughout unless otherwise noted.

\section{Methodology}\label{Methodology}

\subsection{Theoretical Background}\label{theoreticalbackground}

Modelling the optical absorbance, conductivity, or polarizability of low dimensional systems requires an accurate description of their electronic structure, including their electronic band gaps.  As a first-order correction, this requires a rigid upshift of their unoccupied KS eigenenergies.
This can be obtained at an \emph{ab initio} level using the derivative discontinuity correction \cite{derDis} based on the exchange part of the \GLLBsc{} \cite{Gritsenko1995GLLB, Gritsenko1997GLLB2} functional, $\Deltax$. This combines the screening and response parts of the PBEsol \cite{PBEsol} xc potential with a simple orbital-weighted approximation for the exchange part, and can be used to account for the discontinuity of the potential at integer particle numbers\cite{GLLB,Castelli_2012}.  This may be evaluated at the $k$-point corresponding to the band gap using the analytic form\cite{Chlorophyll,PreciadoSWCNTs}
\begin{align}
  \Deltax = \frac{8\sqrt{2}}{3\pi^2} \sum_{n=1}^N \left( \sqrt{\varepsilon_{N+1 }-\varepsilon_{n }}-\sqrt{\varepsilon_{N }-\varepsilon_{n }}\right)\langle \psi_{N+1 }^{} | \frac{\psi_{n }^*\psi_{n }^{}}{\rho}|\psi_{N+1 }\rangle,\label{eqn:Deltax}
\end{align}
where $N$ is the number of electrons, $\psi_n$ is the $n^\textrm{th}$ KS wavefunction, and $\rho$ is the electron density.

An \LCAOTDDFTkomega{} calculation of the dielectric response,  conductivity, or polarizability begins by determining the non-interacting density-density response function\cite{Adler,Wiser} in reciprocal space $k$ and frequency $\omega$ domains 
\begin{align}\label{eqn:chi}
\chi^0_{\bG\bG'} (\bq,\omega) = \frac{1}{\Omega}&
\sum_{\bk}^{\textrm{BZ}} \sum_{n,n'} \frac{w_{\bk}[f(\varepsilon_{n \bk}) - f(\varepsilon_{n' \bk+\bq})]}{\hslash\omega - (\varepsilon_{n' \bk} - \varepsilon_{n \bk+\bq}+\Deltax) +i\eta}\times \nonumber\\
&\langle \psi_{n \bk} | e^{-i (\bq+\bG)\cdot \br} |\psi_{n' \bk+\bq}\rangle
\langle \psi_{n \bk} | e^{i (\bq+\bG')\cdot \br'} |\psi_{n' \bk+\bq}\rangle
\end{align}
where $\bq$ and $\hslash\omega$ are the momentum and energy of the perturbation, $\bG$ and $\bG'$ are reciprocal lattice vectors, $\Omega$ is the unit cell volume, $w_{\bk}$ is the weight of $k$-point $\bk$, 
$f$ is the Fermi-Dirac distribution, $\varepsilon_{n \bk}$ and $\psi_{n \bk} (r)$ are the eigenenergies and eigenfunctions of the $n$th	band at $k$-point $\bk$, $\Deltax$ is the derivative discontinuity correction to the exchange part of the \GLLBsc{} functional from \eqref{eqn:Deltax}, and $\eta \approx 50$~meV is the half width at half maximum of the Lorentzian broadening.

In general, within linear response TDDFT in the random phase approximation (RPA), the dielectric matrix
in reciprocal space 
 is given by
\begin{equation}\label{eqn:dielectric}
\varepsilon_{\bG\bG'}(\bq,\omega) = \delta_{\bG\bG'} - v_{\bG\bG'}(\bq)\chi^0_{\bG\bG'}(\bq,\omega),
\end{equation}
where $\delta_{\bG\bG'}$ is the Kronecker delta and $v_{\bG\bG'}(\bq) = \frac{4\pi}{|\bq+\bG|^2}\delta_{\bG\bG'}$ is the Fourier transform of the Coulomb kernel in 3D\cite{RadialCutoff,Mowbray2014}. The head ($\bG=\bG'=0$) of the dielectric matrix corresponds to the  macroscopic dielectric function in the absence of local crystal field effects (LCFs).

In the so-called optical limit $\|\bq\|\rightarrow 0^+$, the matrix elements corresponding to the head of the dielectric function $\varepsilon_{00}$ in \eqref{eqn:chi} reduce to \cite{TDDFTRPA}

\begin{equation}
\lim_{\bq\rightarrow0}	\langle \psi_{n \bk} | e^{-i \bq\cdot \br} |\psi_{n' \bk+\bq}\rangle = -i\bq\cdot
\frac{\langle \psi_{n \bk} | \nabla |\psi_{n' \bk}\rangle}{\varepsilon_{n' \bk}-\varepsilon_{n \bk} + \Deltax} = |\bq| f_{n n'\bk}^{\hbq},\label{eqn:q0limit}
\end{equation}
where $f_{n n'\bk}^{\hbq}$ is the oscillator strength of the $n \to n'$ transition at $k$-point $\bk$ in the direction of the Bloch vector $\bq$, i.e., $\hbq$.  It is important to note that, in the optical limit and neglecting LCFs, the radial cutoff\cite{RadialCutoff} and zero-padding\cite{Mowbray2014} methods for describing the Coulomb kernel in lower dimensions (0D, 1D, or 2D) reduce to the 3D Coulomb kernel.

In the optical limit the neglect of LCFs has previously been shown effective for gas phase structures and other low-dimensional materials. \cite{Chlorophyll,PreciadoSWCNTs}  This leads to a simplified form for the macroscopic dielectric function in the optical limit

\begin{equation}
\label{eqn:epsilon}
\varepsilon (\hbq,\omega) 
= 1-\frac{4\pi}{\Omega} \sum_{\bk}^{\textrm{BZ}} \sum_{n,n'} \frac{w_{\bk}\left[f(\varepsilon_{n \bk}) - f(\varepsilon_{n' \bk})\right]|f_{n n' \bk}^{\hbq}|^2}{\hslash\omega - (\varepsilon_{n' \bk} - \varepsilon_{n \bk} + \Deltax) +i\eta}.
\end{equation}

Suppressing $k$-point dependence, the matrix elements in \eqref{eqn:epsilon} may be expressed as\cite{GlanzmannTDDFTRPA}
\begin{equation}
\langle\psi_{n}|{\mathbf{\nabla}} |\psi_{n'}\rangle\!=\!\sum_{\mu\nu}\!c_{\nu n}^*c_{\mu n'}^{} \langle \tilde{\phi}_{\nu}|\Proj^\dagger{\mathbf{\nabla}}\Proj|\tilde{\phi}_{\mu}\rangle,\label{matrixelements}
\end{equation}
where $\tilde{\phi}_{\mu}$ are the localized basis functions describing the $n$th KS wave function $|\tilde{\psi}_{n}\rangle = \sum_{\mu}c_{\mu n}|\tilde{\phi}_{\mu}\rangle$ with coefficients $c_{\mu n}$,
and $\Proj$ is the projector augmented wave (PAW) transformation operator\cite{PAW,PAW2,CarstenThesis}
\begin{equation}
\Proj = 1 + \sum_{a i} \left(|\varphi_{i}^a\rangle - |\tilde{\varphi}_{i}^a\rangle\right)\langle \tilde{p}_i^a|,
\end{equation}
where $\tilde{\varphi}_i^a$ and $\varphi_i^a$ are the pseudo and all-electron partial waves for state $i$ on atom $a$ within the PAW formalism,
and $|\tilde{p}_{i}^a\rangle$ are the smooth PAW projector functions.

Since the matrix elements $f_{n n'\bk}^{\hbq}$ must already be calculated to obtain the forces during structural relaxation, calculating the dielectric function using \eqref{eqn:epsilon} simply involves the multiplication of previously calculated matrices.  For this reason, calculations with our \LCAOTDDFTkomega{} code\cite{code} are very efficient, with scaling of $\BigO(NM^2)$ or better\cite{slug}, where $N$ is the number of KS wavefunctions and $M \geq N$ is the total number of basis functions used in the LCAO calculation\cite{PreciadoSWCNTs,Chlorophyll}.  

In order to properly treat low-dimensional materials, we follow approaches based on mean-field theory laid out in Ref.~\citenum{BenedictNanotubes} for non-interacting 1D SWCNTs and in Refs.~\citenum{DobsonLayer} and \citenum{Matthes2016} for non-interacting 2D sheets to compute the polarizability response functions $\alpha(\hbq,\omega)$.  For a given dimension $d$, these may be decomposed into the polarizability for light polarized in periodic directions ``parallel'' to the material, $\hbq_\|$, and in non-periodic directions ``perpendicular'' to the material, $\hbq_\perp$.  These have the general forms
\begin{align}\label{eqn:polarizability}
\alpha_{d} (\hbq_\|, \omega) &= \frac{\Omega_{d}}{4\pi} \left(\varepsilon(\hbq_\|, \omega) -1\right), \\
\alpha_{d} (\hbq_\perp, \omega) &= \frac{\Omega_{d}}{4\pi} \left(1- \frac{1}{\varepsilon(\hbq_\perp,\omega)}\right),
\end{align}
where $\Omega_{d}$ represents the ``cross-section'' of the unit cell for a given dimension $d$, and $\varepsilon(\hbq,\omega)$ is the 3D macroscopic dielectric function obtained from \eqref{eqn:epsilon}. In this way the polarization is defined per molecule or per layer rather than per unit volume.  $\Omega_{\mathrm{0D}}$ is the volume of the unit cell, $\Omega_{\mathrm{1D}}$ is the area of the plane in the unit cell perpendicular to the 1D material, $\Omega_{\mathrm{2D}}$ is the length of the unit cell perpendicular to the plane of the 2D material, and $\Omega_{\mathrm{3D}} = 1$. Using these definitions for the polarizability $\alpha$, we may express the conductivity $\sigma$ for any dimension as
\begin{equation}\label{eqn:conductivity}
\sigma (\hbq, \omega) = -i\omega \alpha(\hbq, \omega).
\end{equation}
This allows us to generalize our \LCAOTDDFTkomega{} code\cite{code} to all classes of materials.
It is important to note, however, that the neglect of LCFs may have important consequences for excitations polarized perpendicular to a given low-dimensional material, \cite{Novko2016} with a spatial averaging taking place over the external field. This makes this \LCAOTDDFTkomega\ method most suitable for computing the axial or in-plane conductivities for 1D and 2D materials, respectively. 


By working in the optical limit $\|\bq\|\rightarrow 0^+$  and neglecting LCFs,  we can define the two-point excitonic density as 
\begin{equation}\label{eqn:twopoint}
\rhoex (\br_e, \br_h, \hbq, \omega) = \frac{4\pi}{\Omega}\sum_{\bk}^{\textrm{BZ}}\sum_{n n'} \frac{\eta^2 w_{\bk} | f_{n n'\bk}^{\hbq}|^2 | \psi_{n\bk} (\br_h)|^2 |\psi_{m\bk} (\br_e)|^2 }{(\hslash\omega-(\varepsilon_{n'\bk}-\varepsilon_{n\bk}+\Deltax))^2+\eta^2},
\end{equation}
where $\br_e$ and $\br_h$ represent the real space locations of the electron and hole, respectively, and $f_{n n'\bk}^{\hbq}$ is the matrix element for the $n\to n'$ transition from \eqref{eqn:q0limit}. 
Based on \eqref{eqn:twopoint}, we may define expressions for the hole and electron densities by averaging over the electron and hole coordinates, respectively, as
\begin{align}
\rho_h (\br_h, \hbq, \omega) &= \int \rhoex (\br_e, \br_h, \hbq, \omega) d\br_e \nonumber\\
&= \frac{4\pi}{\Omega} \sum_{\bk} \sum_{n n'}\frac{\eta^2 w_{\bk} | f_{n n'\bk}^{\hbq}|^2 \left| \psi_{n\bk} (\br_h)\right|^2 }{(\hslash\omega-(\varepsilon_{n'\bk}-\varepsilon_{n\bk}+\Deltax))^2+\eta^2},\label{eqn:rhoh}
\end{align}
\begin{align}
\rho_e (\br_e, \hbq, \omega) &= -\int \rhoex (\br_e, \br_h, \hbq, \omega) d\br_h \nonumber\\
&= -\frac{4\pi}{\Omega} \sum_{\bk} \sum_{n n'}\frac{\eta^2 w_{\bk} | f_{n n'\bk}^{\hbq}|^2 \left| \psi_{n'\bk} (\br_e)\right|^2 }{(\hslash\omega-(\varepsilon_{n'\bk}-\varepsilon_{n\bk} + \Deltax))^2+\eta^2}.\label{eqn:rhoe}
\end{align}

These definitions for the excitonic, electron, and hole densities all satisfy
\begin{equation}
\iint \rhoex (\br_e, \br_h, \hbq, \omega) d\br_h d\br_e = \Imag [\varepsilon (\hbq, \omega)].
\end{equation}
In this way, (\ref{eqn:twopoint}--\ref{eqn:rhoe}) provide means for spatially and energetically resolving the exciton, hole, and electron densities.  
We may then define the electron-hole density difference as\cite{PreciadoSWCNTs}
\begin{align}\label{eqn:rsdelta}
  \Delta \rho (\br, \hbq, \omega) &= \rho_h (\br, \hbq, \omega) + \rho_e (\br, \hbq, \omega)\\
  &=\frac{4\pi}{\Omega} \sum_{\bk} \sum_{n n'}\frac{\eta^2 w_{\bk} | f_{n n'\bk}^{\hbq}|^2 \left(\left| \psi_{n\bk} (\br_h)\right|^2-\left| \psi_{n'\bk} (\br_e)\right|^2\right) }{(\hslash\omega-(\varepsilon_{n'\bk}-\varepsilon_{n\bk} + \Deltax))^2+\eta^2}.\label{eqn:rsdeltarho}
\end{align}

\subsection{Computational Details}

All our DFT calculations employ the PAW method code \textsc{gpaw}\cite{GPAW,GPAWRev} within the atomic simulation environment \textsc{ase}\cite{ASE0,ASE}.  The generalized gradient approximation for solids and surfaces (PBEsol) \cite{PBEsol} was employed throughout for the xc functional.  This allowed a self-consistent calculation of the \GLLBsc{} derivative discontinuity correction\cite{GLLB} to the exchange functional $\Deltax$ from \eqref{eqn:Deltax}. We employed a grid spacing of $h \approx 0.2$~\AA{} and electronic temperature of $\temp \approx 1$~meV with all energies extrapolated to $T\to0$.  The KS wavefunctions have been represented with either LCAOs and a double-$\zeta$-polarized (DZP) basis set\cite{BenchmarkPaper}, after performing convergence tests with basis sets of varying quality up to a quadruple-$\zeta$-polarized (QZP) basis set, or PWs with an converged energy cutoff of $\Ecut \approx 340$~eV.  The radial functions $\zeta(r)$  can describe spatially distinct bonds involving the same atom, whereas polarization\cite{GPAWLCAO} refers to  a mixing with orbitals of higher angular momentum number.  The reliability of \LCAOTDDFTkomega{} is inherently basis set dependent\cite{LCAOBasisSets}, although DZP is an often chosen default for its balance between accuracy and computational efficiency\cite{BenchmarkPaper}.

For C$_{60}$ we relaxed the atomic structure until maximum forces below 0.03~eV/\AA{} were obtained, within a $20 \times 20 \times 20$~\AA{}$^3$ unit cell. Given the 0D nature of C$_{60}$, we performed $\Gamma$-point calculations employing non-periodic boundary conditions, i.e., both the electron density $\rho$ and KS wave functions $\psi_n$ were set to zero at the cell boundaries. 

The (10,0) zigzag and (10,10) armchair SWCNTs' atomic structures were relaxed until maximum forces less than 0.05 eV/\AA{} were obtained, and the unit cells were relaxed parallel to the SWCNT ($z$) axis, yielding unit cell parameters of $L \approx 4.30$ and $2.46$~\AA{}, respectively, including 10~\AA{} of vacuum perpendicular to the nanotube axis. Given the 1D nature of SWNTs, $k$-point samplings of $1\times 1 \times 281$ and $1\times 1 \times 489$ were employed for the (10,0) and (10,10) SWCNTs, respectively, with periodic boundary conditions only along the SWCNT axis, and the electron density and KS wave functions set to zero at the unit cell boundaries perpendicular to the SWCNT's axis. 

For all \Gr{} calculations we have employed an in-plane unit-cell constant of $a = 4.651~a_0 \approx 2.46$~\AA{}, with the \Gr{} layers stacked in a periodic super-lattice along the $z$-axis
separated by a distance of $L = 5a\ \approx 23.255~a_0 \approx 12.3$~\AA{}. To calculate the ground-state electronic density and KS wavefunctions we have employed a dense Monkhorst-Pack $301 \times 301 \times 1$ $k$-point mesh
\cite{MonkhorstPack} over the first BZ with $18$ bands, corresponding to seven unoccupied bands per atom.  This was previously found to be sufficient to converge both the $\pi$ and $\sigma+\pi$
peaks in the energy loss of \Gr{} within RS-TDDFT-RPA \cite{Mowbray2014}. Since our chosen $k$-point mesh does not include the Dirac (K) point, an electronic temperature of $\temp\approx 1$~meV ensures all electronic levels have integer occupations.  A total of 18 bands were employed for all LCAO calculations to accommodate the reduced degrees of freedom present in SZP calculations.

All calculations of \Pn{} have employed the crystal structure found in the \textsc{gpaw} Computational 2D Materials Database\cite{materials-database}. The \Pn{} layers are stacked in a periodic super-lattice along the $z$-axis separated by a distance of $L \approx 17.1$~\AA{}. A dense Monkhorst-Pack $603 \times 603 \times 1$ $k$-point mesh is chosen over the first BZ with $32$ bands, which we found to be sufficient to converge the main peaks up to 20~eV.

For \ATiO, a Monkhorst-Pack $11 \times 11 \times 5$ $k$-point mesh was chosen over the first BZ with approximately 9 unoccupied bands per atom\cite{mowbraytio2}, whereas for \RTiO{} we employed a Monkhorst-Pack $7 \times 7 \times 11$ $k$-point mesh over the first BZ with a similar number of bands\cite{MiganiLong}. Due to the fact that the \LCAOTDDFTkomega{} method is only concerned with the head of the dielectric function, using more bands will only affect the higher energy part of the optical spectra, so using more than the number of bands needed to converge the calculation reasonably is not required. 


It is important to note that the effects of excitonic binding on the spectra have been neglected within the \LCAOTDDFTkomega{} formalism. Such effects may be substantial, especially for low-dimensional systems where electron--hole screening is significantly reduced, e.g., SWCNTs\cite{SpataruPRL2004}.  To determine the impact of excitonic binding on the measured spectra, we have systematic compared our \LCAOTDDFTkomega{} results with $G_0 W_0$-BSE calcultions for C$_{60}$\cite{Despoja_fullerene}, (10,0) SWCNT\cite{PreciadoSWCNTs}, \Pn{}, \ATiO{}\cite{mowbraytio2}, and \RTiO{}\cite{MiganiLong}.

Our $G_0 W_0$-BSE calculations for the (10,0) SWCNT and \Pn{} were performed using the PW implementation within \textsc{gpaw}\cite{Huser_2013}, employing reduced $1\times 1 \times 64$ and $33 \times 47 \times 1$ samplings of the BZ, respectively.  We used the Godby-Needs plasmon-pole approximation\cite{PlasmonPole,PlasmonPole1995,PlasmonPoleModels} to describe the screening $W$ and 1D and 2D truncation schemes for the Coulomb kernel\cite{RadialCutoff} to remove spurious interactions with periodic images orthogonal to the SWCNT's axis and the \Pn{} sheet, respectively.

One method for representing the electron and hole densities is simply, for a given energy, to present a 3D isosurface plot for both the electron and hole densities or their difference. Another option is to project the 3D voxel data onto an axis within the unit cell of a given material.  This projection can be accomplished by summing up voxels along planes in the unit cell perpendicular to the desired axis.  The functionality for producing both types of spatial representations of the electron and hole densities has been incorporated into our \LCAOTDDFTkomega{} code\cite{code}.

The axes of electron and hole density projections cannot be chosen arbitrarily within our \LCAOTDDFTkomega{} code\cite{code}, as the summation method for projection requires that the axis passes through voxels in a co-linear way. Beyond this requirement, projection of the spatially-resolved electron and hole densities is not limited to any specific crystal structure so long as it has dimension greater than zero. 

\section{Results and Discussion}\label{results}


\subsection{0D Fullerene}

We begin our assessment of the \LCAOTDDFTkomega{} method by considering the non-periodic or 0D carbon-based system of an isolated fullerene (C$_{\text{60}}$) molecule.  C$_{\text{60}}$ provides an excellent system for benchmarking optical absorption spectra due to both its usefulness in OPVs\cite{AngelFullerenesPhotovoltaics,PolymerFullerenePhotovoltaic,FullereneOnlyPhotovoltaic,Fullerene_apl1} and the availability of experimental measurements\cite{Berkowitz_1999,Hare_2013} and theoretical calculations\cite{Despoja_2014,Despoja_fullerene}.

In Table~\ref{tbl:fullerene} 
\begin{table}
	\caption{Measured and calculated energies of the third bright $\pi - \pi^*$ exciton of fullerene (C$_{60}$) $\hslash\omega_{\pi}$  in eV.}\label{tbl:fullerene}
  \begin{ruledtabular}
	\begin{tabular}{lc}
	  Method & $\hslash\omega_{\pi}$ (eV)\\\hline
		Measurement in Gas Phase & $5.97$\footnote{Ref.~\citenum{Berkowitz_1999}.}\\
		Measurement in Hexane & $5.86$\footnote{Ref.~\citenum{Hare_2013}.} \\
		$G_0 W_0$-BSE (VASP) & $6.04$\footnote{Ref.~\citenum{Despoja_fullerene}.}\\
		$G_0 W_0$-BSE (QE) & $6.50$\footnote{Ref.~\citenum{Despoja_2014}.}\\
		\LCAOTDDFTkomega{} ($\varepsilon_{n'}-\varepsilon_n$)& $4.79$\footnote[5]{This work.} \\
		\LCAOTDDFTkomega{} ($\varepsilon_{n'}-\varepsilon_n + \Deltax$) & $5.56$\footnotemark[5] \\
	\end{tabular}
  \end{ruledtabular}
\end{table}
we directly compare the energies obtained for C$_{60}$'s third bright $\pi - \pi^*$ exciton, $\hslash\omega_\pi$, from experiments in gas phase\cite{Berkowitz_1999} and hexane solution\cite{Hare_2013} with $G_0W_0$-BSE\cite{Despoja_fullerene,Despoja_2014} and our \LCAOTDDFTkomega{} calculations neglecting and including the derivative discontinuity correction $\Deltax$.  While both $G_0W_0$-BSE and \LCAOTDDFTkomega{} including $\Deltax$ reproduce the measured excitonic energy semi-quantitatively, neglecting the derivative discontinuity correction leads to an underestimation by more than 1~eV. This clearly demonstrates the essential role played by the derivative discontinuity correction $\Deltax$ in providing a sufficiently accurate description of a system's electronic structure to reproduce the measured spectra.  

\begin{figure}
	\includegraphics[width=\columnwidth]{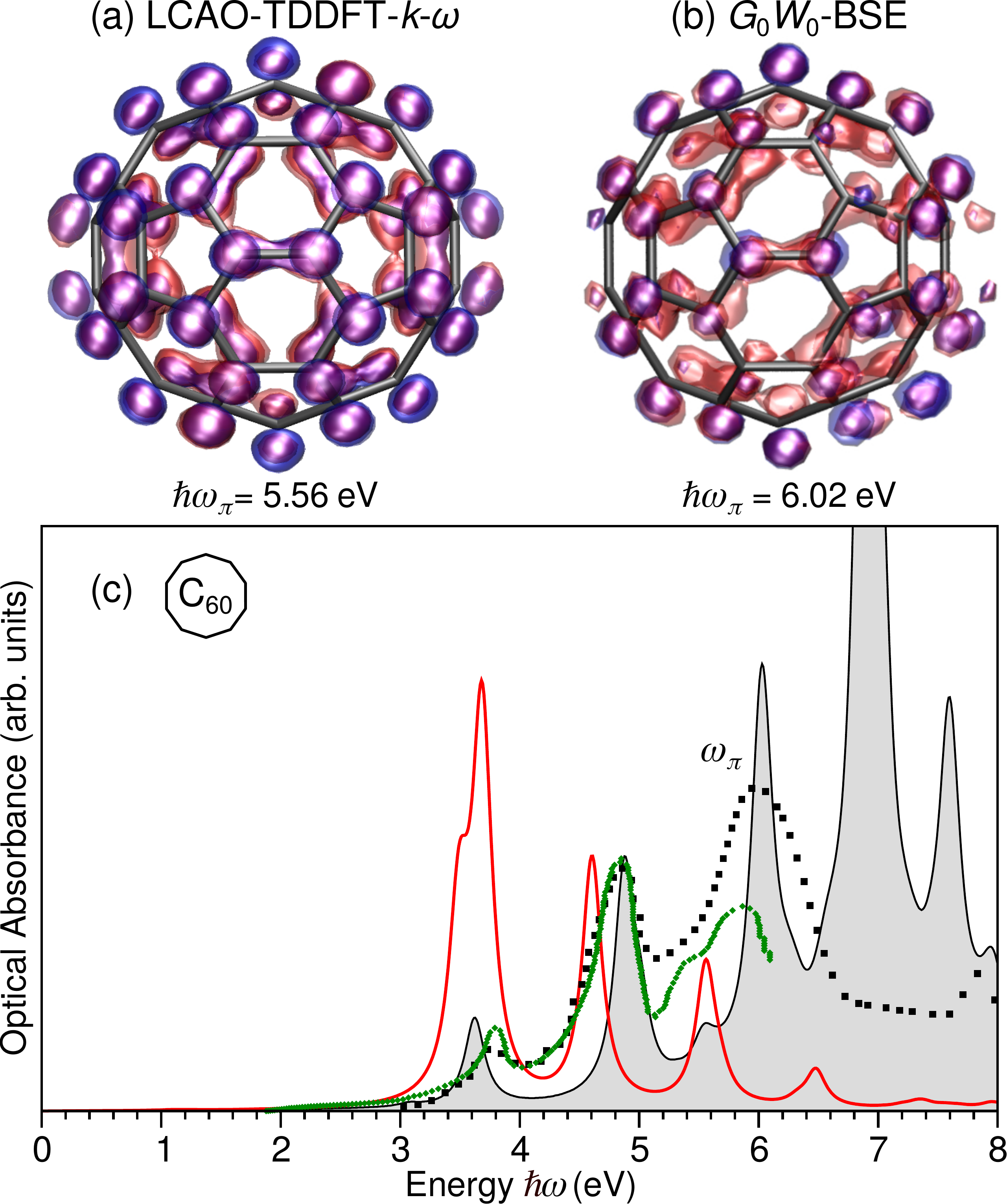}
	\caption{Fullerene's (C$_{60}$) electron (blue) and hole (red) densities for the third bright $\pi - \pi^*$  exciton $\omega_\pi$ from (a) \LCAOTDDFTkomega{} and (b) $G_0W_0$-BSE\cite{Despoja_fullerene} and (c) optical absorbance spectra from \LCAOTDDFTkomega{} (red thick solid line) and $G_0W_0$-BSE (black thin solid line)\cite{Despoja_fullerene} calculations and measurements in gas phase (black squares)\cite{Berkowitz_1999} and hexane (green diamonds)\cite{Hare_2013}.}\label{plot-fullerene}
\end{figure}

In Figure \ref{plot-fullerene}(a) and (b) we show the spatial distribution of the electron and hole densities associated with C$_{60}$'s third bright $\pi - \pi^*$ exciton, $\rho_e(\rv_e,\omega_\pi)$ and $\rho_h(\rv_h,\omega_\pi)$, respectively, from \LCAOTDDFTkomega{} at $\hslash\omega_\pi \approx 5.6$~eV and $G_0 W_0$-BSE\cite{Despoja_fullerene} at $\hslash\omega_\pi \approx 6.0$~eV.  In both cases, the electron density $\rho_e(\br,\omega_\pi)$ is predominately on $\pi$ anti-bonding orbitals located on the exterior of the C$_{60}$, whereas the hole density $\rho_h(\br,\omega_\pi)$ is predominately on $\pi$ bonding orbitals located on the interior of the C$_{60}$, as expected\cite{Despoja_fullerene}.  Overall, the two methods yield electron and hole densities in semi-quantitative agreement, justifying the use of our \LCAOTDDFTkomega{} code\cite{code}. 

In Figure \ref{plot-fullerene}(c), we compare experimental photoexcitation data\cite{Berkowitz_1999,Hare_2013}, \textit{ab initio} models utilizing the quasiparticle $G_0 W_0$ eigenenergies within the Bethe--Salpeter equation ($G_0W_0$-BSE) to better describe excitonic effects, and our \LCAOTDDFTkomega{} calculations including the derivative discontinuity correction to the \GLLBsc{} exchange functional $\Deltax$. We find that both $G_0 W_0$-BSE and our \LCAOTDDFTkomega{} calculations yield well-separated single-transition peaks at energies in semi-quantitative agreement with the experimental measurements in solution\cite{Berkowitz_1999,Hare_2013}.

To allow a direct comparison of the measured and calculated spectra, in Figure~\ref{plot-fullerene}(c) we have normalized the peak intensities by C$_{60}$'s second $\pi - \pi^*$ excitation at about 5~eV.  Whereas $G_0 W_0$-BSE yields peaks gradually increasing in intensity in qualitative agreement with the experimental spectra, \LCAOTDDFTkomega{} shows the opposite behavior.  This may be related to \LCAOTDDFTkomega{}'s neglect of charge-transfer excitations or other excitonic effects which are better described at the $G_0W_0$-BSE level.  However, the $G_0W_0$-BSE spectrum's most intense peak at about 7~eV is almost completely absent in both the \LCAOTDDFTkomega{} and experimental\cite{Berkowitz_1999} spectra.

It is also worth noting that the inclusion of the \GLLBsc{} correction from \eqref{eqn:Deltax} not only blue-shifts the calculated spectrum, but also reduces the intensities of the lower energy peaks, in better agreement with experiment and $G_0W_0$-BSE calculations.  This is clearly seen from the inclusion of $\Deltax$ in the denominator of the matrix elements $f_{n n'\bk}^{\hbq}$ in \eqref{eqn:q0limit}.  

Despite these differences, it is apparent that the computationally less intensive \LCAOTDDFTkomega{} method is able to capture both the spatial distribution, peak locations, and peak intensities qualitatively and semi-quantitatively for this prototypical 0D system. We note that the absorption data for C$_{60}$ in hexane solution was not adjusted by a Chako factor \cite{Berkowitz_1999} since we are mostly interested in the location of the peaks.  Having demonstrated the reliability of the \LCAOTDDFTkomega{} method for describing both the optical absorption spectra and electron and hole density distributions for a non-periodic 0D carbon-based molecular system, we shall next consider extended semi-periodic systems.

\subsection{1D (10,10) \& (10,0) SWCNTs}

To continue our assessment of the \LCAOTDDFTkomega{} method, we next consider a structure that is periodic in only one direction or 1D.  Specifically, we consider the prototypical 1D carbon-based system of an isolated single-walled carbon nanotube (SWCNT).  SWCNTs provide excellent systems to benchmark optical absorption spectra due to their utility as conducting layers in OPVs\cite{Fullerene_apl1,berson2007elaboration,kymakis2002single,bartelmess2010phthalocyanine}, the tailorability of their optical properties\cite{baughman2002carbon}, and the availability of experimental optical absorption\cite{OpticalAbsorbance} and electron energy loss spectra (EELS)\cite{SWCNTEELS}.

In fact, for a set of fifteen semiconducting and four metallic SWCNTs, our \LCAOTDDFTkomega{} code\cite{code} with the $\Deltax$ correction has already been shown to reproduce these experimental measurements semi-quantitatively\cite{PreciadoSWCNTs}.  For this reason we shall focus herein on the axial optical conductivity $\sigma(\hbq_\|, \omega)$ obtained from the 1D polarizability $\alpha_{\text{1D}}(\hbq_\|,\omega)$ using \eqref{eqn:polarizability} and \eqref{eqn:conductivity}.  Specifically, we consider the conductivity of two prototypical SWCNTs: the metallic armchair (10,10) SWCNT and the semiconducting zigzag (10,0) SWCNT.

\begin{figure}
  \includegraphics[width=\columnwidth]{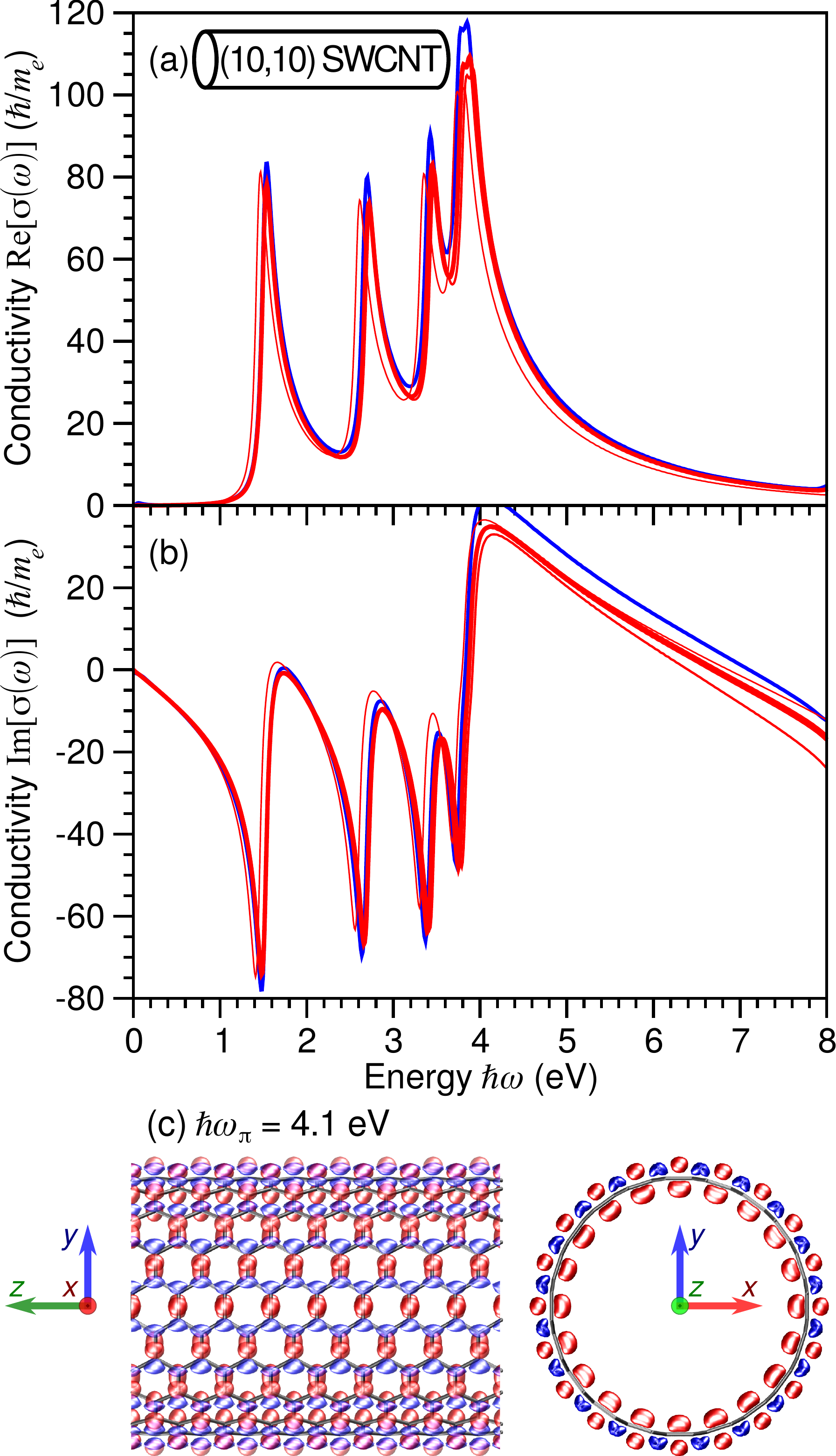}
  \caption{Metallic (10,10) SWCNT axial (a) real and (b) imaginary parts of the conductivity $\sigma(\hbq_\|, \omega)$ in $\hslash/m_e$ versus energy $\hslash\omega$ in eV for axially-polarized light from \PWTDDFTkomega{} (blue lines) and \LCAOTDDFTkomega{} (red lines) with single (SZP), double (DZP), triple (TZP), and quadruple (QZP) $\zeta$-polarized basis sets (in order of increasing thickness) and (c) positive (red) and negative (blue) isosurfaces for the $\pi$ plasmon's electron-hole density difference $\Delta\rho (\br, \hbq_\|,\omega_\pi) = \rho_e (\br, \hbq_\|, \omega_\pi) + \rho_h (\br, \hbq_\|, \omega_\pi)$ at $\hbar\omega_\pi \approx 4.1$~eV with axes shown as insets.}\label{fig:1010SWCNT}
\end{figure}

In Figure~\ref{fig:1010SWCNT} we compare the axial conductivity of the metallic (10,10) SWCNT from \PWTDDFTkomega{} and \LCAOTDDFTkomega{} employing a systematic improvement from single $\zeta$ polarized (SZP) to quadruple $\zeta$ polarized (QZP) LCAO basis sets.  While the \PWTDDFTkomega{} conductivity is somewhat greater than that obtained with \LCAOTDDFTkomega{}, irrespective of the basis set employed, the two methods are in semi-quantitative agreement.  This underestimation of intensities by \LCAOTDDFTkomega{} could be partially attributed to an improved description of unoccupied wave functions when employing a PW representation.  Comparing the \LCAOTDDFTkomega{} spectra employing various levels of basis sets, we find a DZP basis set is already sufficient to converge the conductivity spectra.    

In each case, the real part of the (10,10) SWCNT's conductivity $\Real[\sigma(\hbq_\|, \omega)]$ consists of a series of intense peaks at 1.5, 2.7, 3.4, and 3.9~eV, of $\sim 100~\hslash/m_e$ per SWCNT. This intensity is consistent with the ballistic conductance expected for a metallic armchair SWCNT.

Figure~\ref{fig:1010SWCNT}(c) shows the spatial distribution of the (10,10) SWCNT's $\pi - \pi^*$ electron-hole density difference $\Delta\rho (\br, \hbq_\|, \omega_\pi)$ at $\hslash\omega_\pi \approx 4.1$~eV from \eqref{eqn:rsdeltarho}.  This exciton's hole is located primarily on $\pi$ orbitals of the C--C bonds wrapping around the (10,10) SWCNT, whereas the excited electron is primarily located above the C--C bond along the (10,10) SWCNT's axis.  We also clearly see the hole has greater weight inside the (10,10) SWCNT whereas the excited electron is predominantly outside the (10,10) SWCNT.   Overall, the electron density  is arranged in a series of ``strips'' running along the outside of the nanotube parallel to its axis, whereas the hole is more localized.

It is important to note that, since the (10,10) armchair SWCNT is metallic, the derivative discontinuity correction $\Deltax$ may be taken implicitly to be zero.  By considering the spectra of the semiconducting (10,0) zigzag SWCNT, we may probe the relevance of the derivative discontinuity correction to the description of 1D systems such as SWCNTs.

\begin{figure}
  \includegraphics[width=\columnwidth]{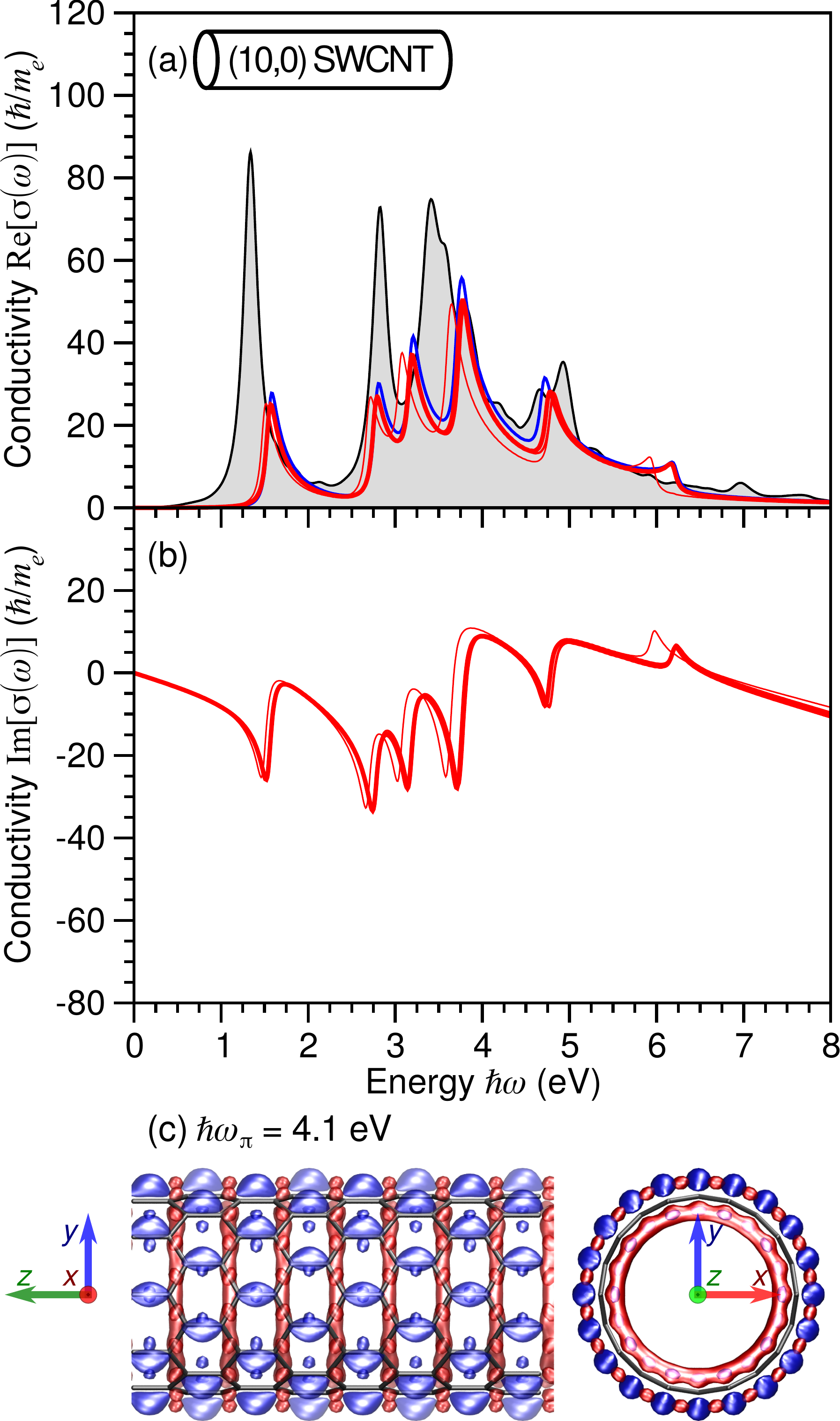}
  \caption{Semiconducting (10,0) SWCNT axial (a) real and (b) imaginary parts of the conductivity $\sigma(\hbq_\|, \omega)$ in $\hslash/m_e$ versus energy $\hslash\omega$ in eV for axially-polarized light from $G_0W_0$-BSE (black), \PWTDDFTkomega{} (blue), and \LCAOTDDFTkomega{} (red) with single (SZP), double (DZP), triple (TZP), and quadruple (QZP) $\zeta$-polarized basis sets (in order of increasing thickness) and (c) positive (red) and negative (blue) isosurfaces for the $\pi$ plasmon's electron-hole density difference $\Delta\rho (\br, \hbq_\|, \omega_\pi) = \rho_e (\br, \hbq_\|, \omega_\pi) + \rho_h (\br, \hbq_\|, \omega_\pi)$ at $\hbar\omega_\pi \approx 4.1$~eV with axes shown as insets.}\label{fig:100SWCNT}
\end{figure}

In Figure~\ref{fig:100SWCNT} we compare the axial conductivity of the semiconducting (10,0) armchair SWCNT from \PWTDDFTkomega{}, $G_0W_0$-BSE, and \LCAOTDDFTkomega{}, again employing a systematic improvement from SZP to QZP LCAO basis sets. As was the case for the metallic (10,10) SWCNT, we find our \LCAOTDDFTkomega{} calculations somewhat underestimate the intensities of the \PWTDDFTkomega{} calculations, irrespective of the basis set employed, whereas the two methods yield semi-quantitative agreement in both their intensities and peak positions. Our $G_0W_0$-BSE calculations\cite{PreciadoSWCNTs} show the effects of excitonic binding on the axial conductivity, which is expected to be quite strong for semiconducting SWCNTs.  $G_0W_0$ yields a quasiparticle band gap of $\EgapQP{} \approx 1.83$~eV, consistent with $\EgapQP{} \approx 1.72$~eV from Ref.~\citenum{GWSWNT}, while BSE yields a strong binding energy for the first bright exciton of $\Ebind = \EgapQP - \hslash\omega_{\textit{ex}} \approx 0.49$~eV.  Altogether, we obtain a $G_0W_0$-BSE energy for the first transition of $E_{11} \approx 1.34$~eV, in semi-qualitative agreement with $E_{11}\approx 1.55$~eV from our \LCAOTDDFTkomega{} and \PWTDDFTkomega{} calculations. Our $G_0W_0$-BSE calculations are in qualitative agreement with our \LCAOTDDFTkomega{} results for the axial conductivity intensity, with $G_0W_0$-BSE yielding somewhat greater intensities for the $E_{11}$ and $E_{22}$ transitions.  This is consistent with the inclusion of excitonic effects at the BSE level.  

For the (10,0) zigzag SWCNT, the application of the derivative discontinuity correction $\Deltax$ changes the calculated spectra qualitatively, not only in the peak positions, but also the intensities.  For the (10,0) SWCNT we obtain a significant derivative discontinuity correction of $\Deltax \approx 0.788$~eV, which is half the energy of the first intense peak in the conductivity spectra, at $\hslash\omega \approx 1.55$~eV.  However, since energy shifts are only employed \emph{a posteriori} by the \PWTDDFTkomega{} implementation within \textsc{gpaw}, it is necessary to include the renormalization of the matrix elements in \eqref{matrixelements} in an \emph{ad hoc} manner when calculating the conductivity from the 3D dielectric function $\varepsilon(\omega)$ obtained from \PWTDDFTkomega{} (see Appendix \ref{section:ddc}).  Although this approximate renormalization works reasonably well for the real part of the conductivity shown in Figure~\ref{fig:100SWCNT}(a), it fails when applied to the imaginary part of the conductivity. For this reason, we do not include \PWTDDFTkomega{} results in Figure~\ref{fig:100SWCNT}(b).  Overall, the real part of the semiconducting (10,0) SWCNT's conductivity $\Real[\sigma(\hbq_\|, \omega)]$ consists of a series of intense peaks at 1.6, 2.8, 3.2, 3.7, and 4.8, of $\sim 40~\hslash/m_e$ per SWCNT, qualitatively half that of the metallic (10,10) SWCNT.  

Figure~\ref{fig:100SWCNT}(c) shows the spatial distribution of the (10,0) SWCNT's $\pi - \pi^*$ electron-hole density difference $\Delta\rho (\br, \hbq_\|, \omega_\pi)$ at $\hslash\omega_\pi \approx 4.1$~eV from \eqref{eqn:rsdeltarho}.   In this case, the exciton's hole is again located primarily on $\pi$ orbitals of the C--C bonds wrapping around inside the (10,10) SWCNT, whereas the excited electron is primarily located above the C--C bonds along the (10,0) SWCNT's axis.  We also again see the hole has greater weight inside the (10,0) SWCNT whereas the excited electron is predominantly outside the (10,0) SWCNT.  However, unlike the (10,10) SWCNT, it is the hole density which is arranged in a series of ``rings'' inside the (10,0) SWCNT along its circumference, whereas the excited electron is more localized.

For both metallic and semiconducting SWCNTs, we find the hole is located predominantly on the inside whereas the excited electron is located predominantly on the outside of the tube.  This is consistent with both the \LCAOTDDFTkomega{} and $G_0W_0$-BSE spatial distributions for C$_{60}$'s third bright exciton (\emph{cf.}~Figure~\ref{plot-fullerene}(a,b)).  Interestingly, we also find in both the (10,10) and (10,0) SWCNTs that the hole's density is arranged predominantly ``around'' the nanotube's circumference, whereas the excited electron's density is arranged predominantly ``along'' the nanotube's axis.  It is when these densities are aligned with the zigzag direction on the SWCNT that they become spatially delocalized in rings inside or strips outside the SWCNT, respectively.  To probe the generality of these observations, we will next consider carbon's 2D allotrope, graphene.

\subsection{2D Graphene \& Phosphorene}

We will now assess the reliability of the \LCAOTDDFTkomega{} method for describing isolated surface layers that are non-periodic in only one direction, or 2D.  We will first consider the prototypical 2D system of an isolated 2D graphene layer (\Gr).  \Gr{} is an excellent material for benchmarking due to its wide range of technologically relevant optoelectronic properties\cite{Vakil_2011,Pai-Yen2017}, its place as a ``toy'' system due to its simplicity from a computational point of view\cite{Susi_2015}, and the great breadth of literature studying this material\cite{Geim_2007}.  

\begin{figure}
	\includegraphics[width=\columnwidth]{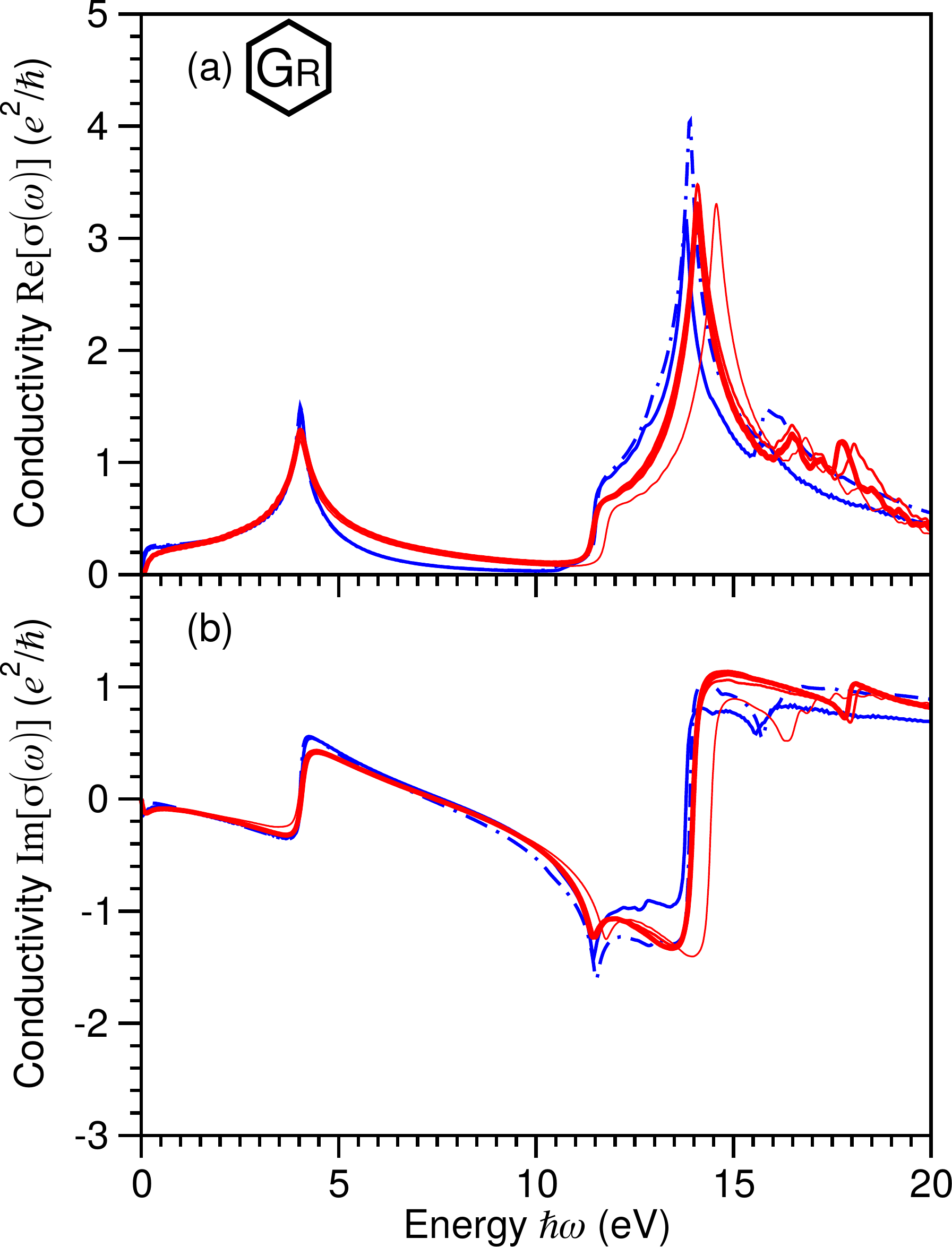}
	\caption{Graphene (\Gr) in-plane (a) real and (b) imaginary parts of the conductivity $\sigma(\hbq_\|, \omega)$ in $e^2/\hslash$ versus energy $\hslash\omega$ in eV from \PWTDDFTkomega{} (blue lines) with \textsc{qe} (dash-dotted) and \textsc{gpaw} (solid) and \LCAOTDDFTkomega{} (red solid lines) with single (SZP), double (DZP), triple (TZP), and quadruple (QZP) $\zeta$-polarized basis sets (in order of increasing thickness). }\label{plot:lcao}
\end{figure}
In Figure \ref{plot:lcao} we compare \Gr's  in-plane conductivity $\sigma(\omega)$ from \PWTDDFTkomega{}  using Quantum Espresso (\textsc{qe}) or \textsc{gpaw} to \LCAOTDDFTkomega{} calculations employing either SZP, DZP, TZP, or QZP basis sets.  Our results in Figure \ref{plot:lcao} show excellent agreement between the \PWTDDFTkomega{} and \LCAOTDDFTkomega{} spectra up to the $\pi \rightarrow \pi^*$ transition near $4.2$~eV, after which \LCAOTDDFTkomega{} predicts a much longer-ranged tail than \PWTDDFTkomega{}. The location of the $\sigma \rightarrow \sigma^*$ transition for \LCAOTDDFTkomega{} employing DZP, TZP, or QZP basis sets also agree with that predicted by \PWTDDFTkomega{}. Finally, \Gr's conductivity exhibits additional peaks beyond the $\sigma \rightarrow \sigma^*$ transition. A small peak near $16$~eV is evident in both \PWTDDFTkomega{} and \LCAOTDDFTkomega{} spectra, whereas \LCAOTDDFTkomega{} calculations with DZP, TZP, or QZP basis sets have a secondary peak near $18$~eV. To determine whether this peak has a  physical origin or is an artifact of the LCAO basis set, we must consider the spatial distribution of the exciton. Overall, the choice of the DZP basis set appears to be sufficient to capture the main peak locations and intensities for the conductivity of \Gr. 

So far we have restricted consideration to carbon-based low-dimensional materials. To further test the range of applicability of the \LCAOTDDFTkomega{} method, we will now consider another prototypical 2D material, monolayer phosphorene (\Pn).  This material has found many potential uses in electronic and photonic applications due to its mid-infrared band gap \cite{Li_2016}, strong layer dependence, and its ability to sustain hyperbolic plasmons due to the anisotropy present in the dielectric tensor \cite{phosphorene}.

\begin{figure}
	\includegraphics[width=\columnwidth]{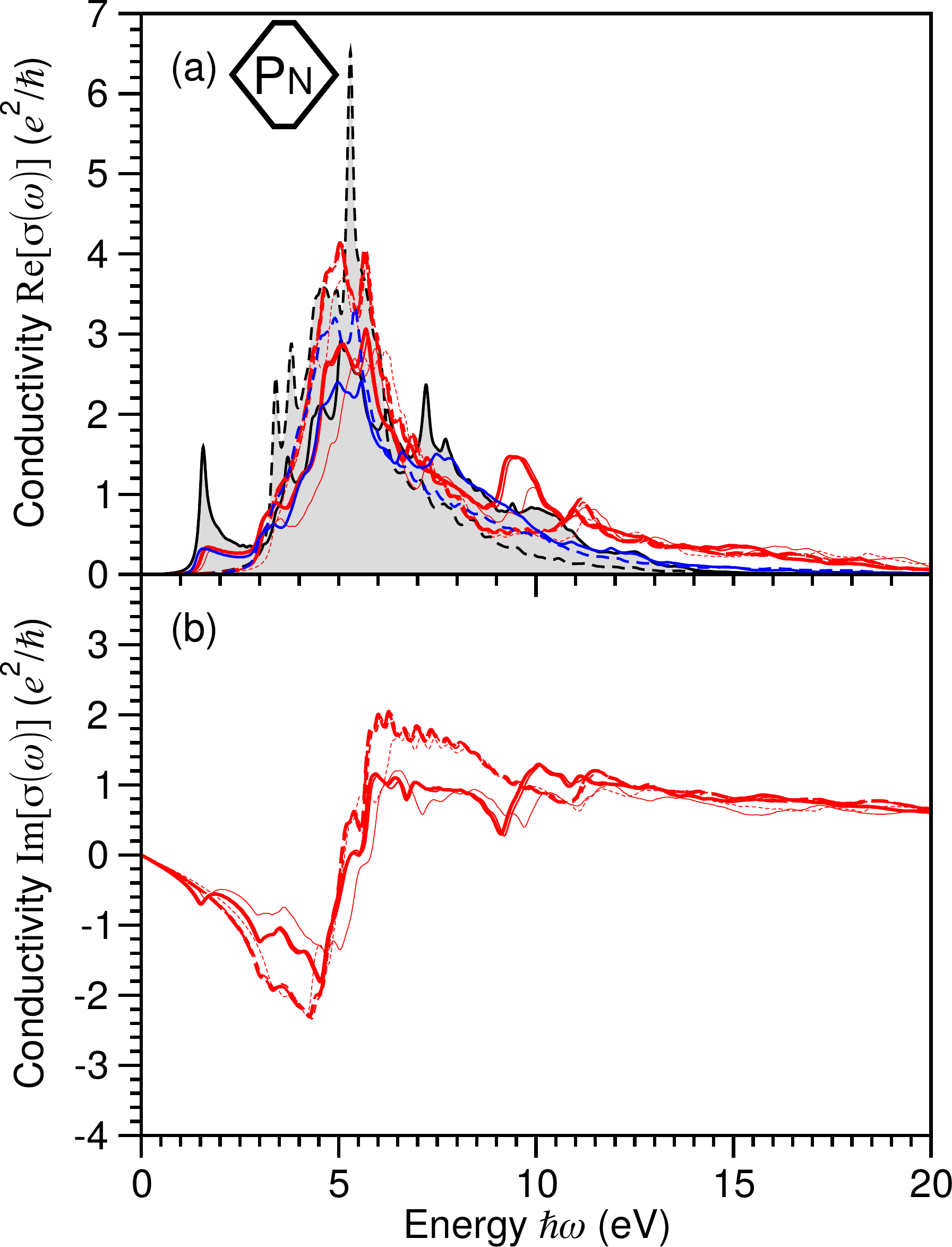}
	\caption{Phosphorene (\Pn) in-plane (a) real and (b) imaginary parts of the conductivity $\sigma(\hbq_\|, \omega)$ in $e^2/\hslash$ versus energy $\hslash\omega$ in eV from $G_0 W_0$-BSE (black), \PWTDDFTkomega{} (blue) and \LCAOTDDFTkomega{} (red) for $\hbq_x$ (solid lines) and $\hbq_y$ (dashed lines) polarized light.}\label{plot:phosphorene_conductivity}
\end{figure}

In Figure \ref{plot:phosphorene_conductivity} we show \Pn's \PWTDDFTkomega{}, $G_0W_0$-BSE, and \LCAOTDDFTkomega{} in-plane conductivity. Due to \Pn's well-known in-plane anisotropy\cite{2Dplasmons, phosphorene}, conductivities are shown in both the $x$ and $y$ directions in Figure \ref{plot:phosphorene_conductivity}.  \PWTDDFTkomega{}, $G_0W_0$-BSE, and \LCAOTDDFTkomega{} all capture the anisotropic peak in the conductivity near 1.8~eV in the $x$-direction, and the greater intensity of the main peak in the conductivity near 5~eV in the $y$-direction.

Overall, we see that both $G_0W_0$-BSE and \LCAOTDDFTkomega{} yield greater conductivities for \Pn relative to \PWTDDFTkomega{} for energies beyond 3~eV. While the magnitude of the matrix elements $\langle \psi_n | \nabla | \psi_{n'}\rangle$  in \eqref{eqn:q0limit} are generally smaller when employing an LCAO representation of the KS orbitals \cite{LCAOBasisSets} due to difficulties representing unoccupied wave functions, this is not always the case. For the conductivity at higher energies \LCAOTDDFTkomega{} yields longer tails in both \Gr{} and \Pn{} relative to \PWTDDFTkomega{} ang $G_0W_0$-BSE, showing another main difference between the three approaches.

\begin{figure*}
	\includegraphics[width=\textwidth]{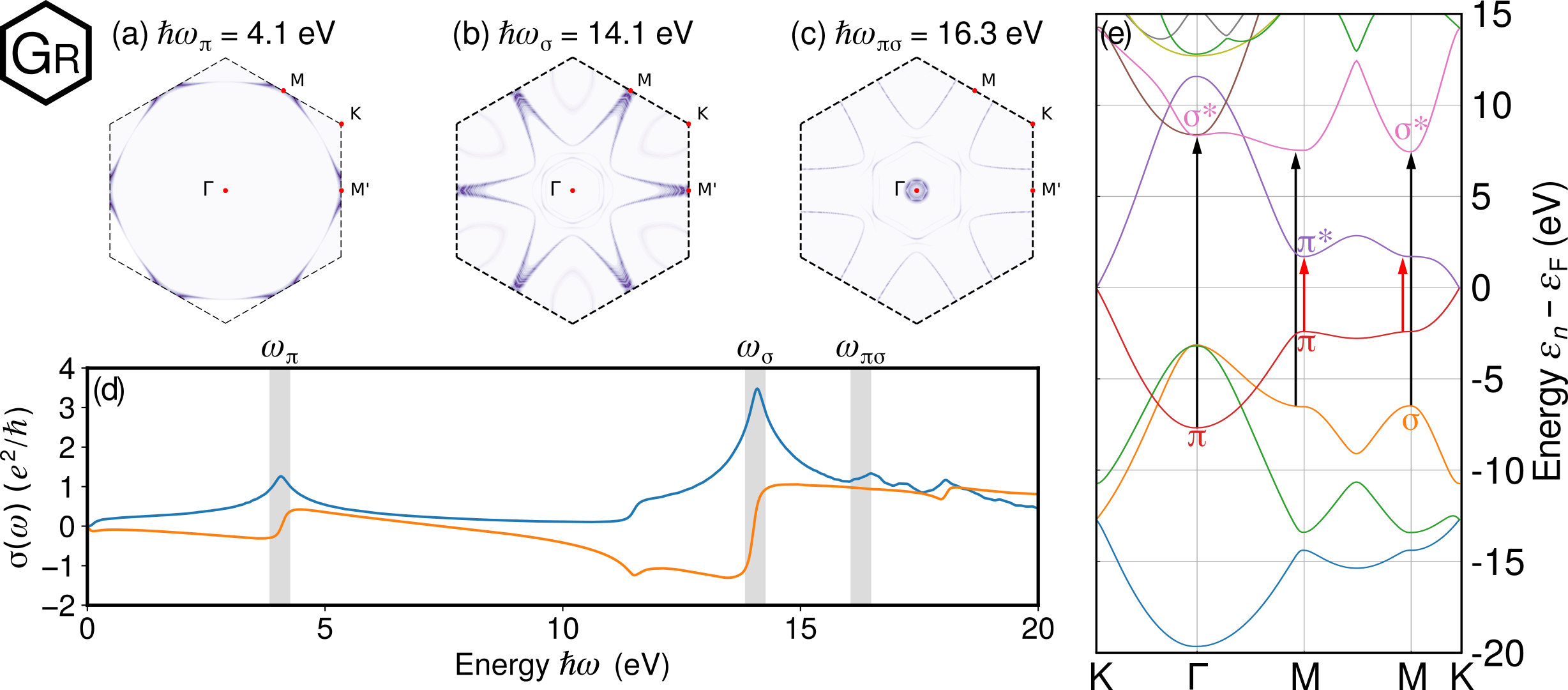}
	\caption{Graphene's (\Gr) (a) $\pi -\pi^*$ transition $\hslash\omega_\pi \approx 4.1$~eV, (b) $\sigma - \sigma^*$ transition $\hslash\omega_\sigma \approx 14.1$~eV, and (c) $\pi-\sigma^*$ transition $\hslash\omega_{\pi\sigma*} \approx 16.3$~eV weights in reciprocal $k$-space, in-plane (d) real (blue) and imaginary (red) parts of the conductivity $\sigma(\hbq_\|, \omega)$ in $e^2/\hslash$ versus energy $\hslash\omega$ in eV from \LCAOTDDFTkomega{}, and (e) band structure.}\label{plot:graphene_bz}
\end{figure*}

\begin{figure*}
  \includegraphics[width=\textwidth]{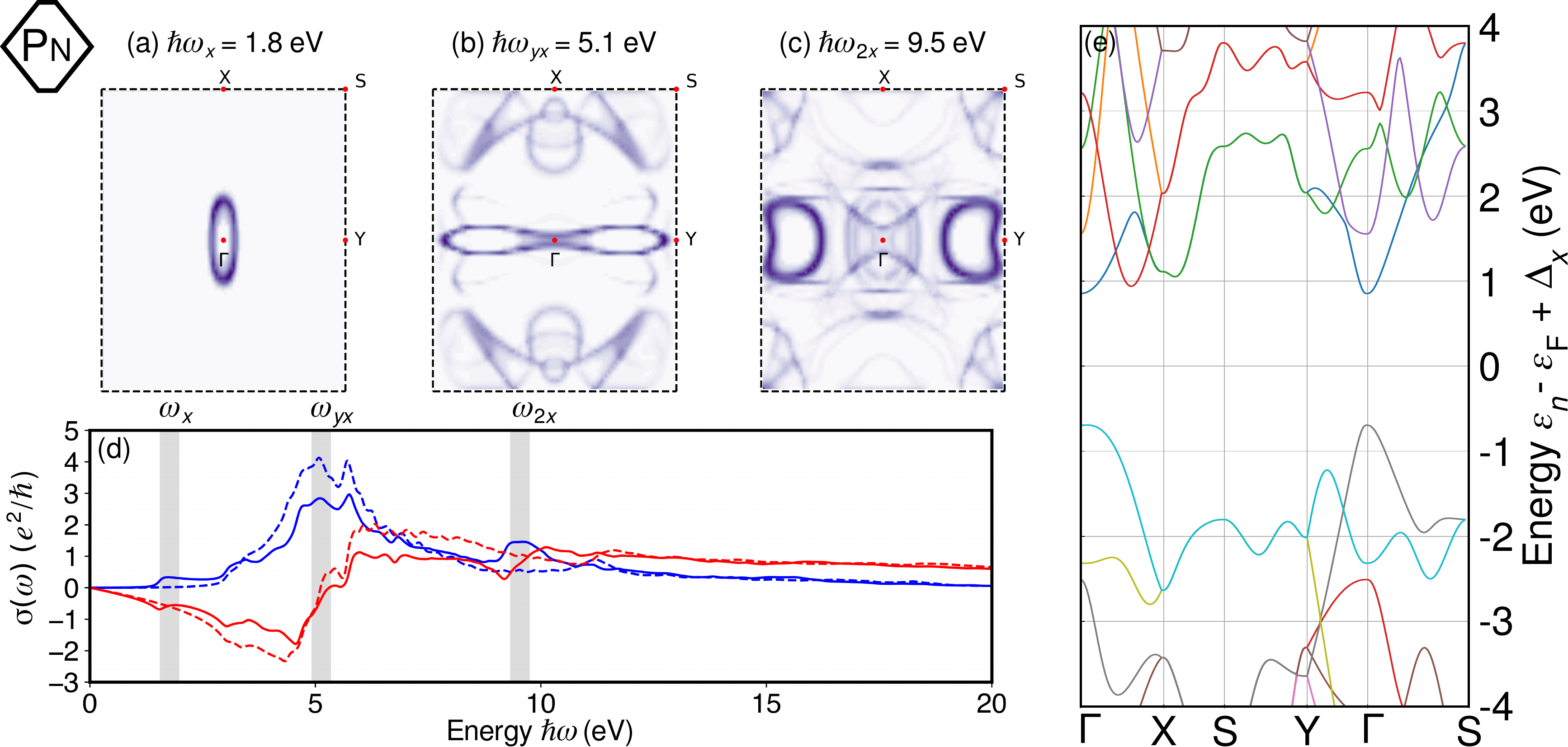}
\caption{Phosphorene's (\Pn) (a) $x$-polarized transition $\hslash\omega_x \approx 1.8$~eV , (b) $y$ and $x$-polarized transition $\hslash\omega_{yx} \approx 5.1$~eV, and (c) second $x$-polarized transition $\omega_{2x} \approx 9.5$~eV weights in reciprocal $k$-space, in-plane (d) real (blue) and imaginary (red) parts of the conductivity $\sigma(\hbq_\|, \omega)$ in $e^2/\hslash$ versus energy $\hslash\omega$ in eV from \LCAOTDDFTkomega{} for $\hbq_x$ (solid lines) and $\hbq_y$ (dashed lines) polarized light, and (e) band structure.}\label{plot:phosphorene_bz}
\end{figure*}

Since we restrict consideration to the head of the dielectric function $\varepsilon_{00}(\omega)$ in \eqref{eqn:dielectric}, thereby neglecting LCFs, the \LCAOTDDFTkomega{} method provides direct access to the individual transitions.  In this way, we may visualize the exciton's distribution within the BZ via the oscillator strengths $f_{n n'\bk}^{\hbq}$ defined in \eqref{eqn:q0limit}, which are directly related to the dipole transition matrix elements\cite{Adler}. In Figures~\ref{plot:graphene_bz} and \ref{plot:phosphorene_bz} we show, for a given $k$-point and energy $\hslash \omega$, the sum of oscillator strengths $f_{n n'\bk}^{\hbq_x}$ for transitions $n \to n'$ within  $\hslash \Delta \omega = \pm 0.2$~eV of $\hslash\omega$ weighted by a Lorentzian broadening of $0.1$~eV. Despite the choice to sum oscillator strengths $f_{n n'\bk}^{\hbq}$ for excitations polarized in-plane, these diagrams can also be made to visualize the relevant bands for excitations in other directions. Since this can be done for any $k$-point, a contour map over the entire BZ can be constructed for a given energy and associated range and broadening parameters. This yields more information at a given energy than the band structure by also showing connections within the BZ beyond linear cross-sections between different high-symmetry points.

In Figure~\ref{plot:graphene_bz}(a), (b), and (c) we show the weights of \Gr{}'s transitions over its hexagonal BZ for its $\pi-\pi^*$, $\sigma - \sigma^*$, and $\pi - \sigma^*$ excitations, respectively. Due to the symmetry of \Gr's crystal structure, these transitions must exhibit a six-fold symmetry over the BZ. Figure~\ref{plot:graphene_bz}(a) and (b) show that for the conductivity peaks located at $4.1$~eV and $14.1$~eV, respectively, the bulk of the transitions occur at or near the high symmetry M-point, with the $\pi-\pi^*$ transition distributed along the M$\to$M' direction\cite{Mowbray2014}. This may also be seen from the band structure of Figure~\ref{plot:graphene_bz}(e), where the $\pi - \pi^*$ and $\sigma - \sigma^*$ transitions at the M-point are marked using arrows. Likewise, the peak at $16.3$~eV in Figure~\ref{plot:graphene_bz}(c) can be attributed to a $\pi - \sigma^*$ transition occurring predominantly at the $\Gamma$ point (\emph{cf.}\ Figure~\ref{plot:graphene_bz}(e)). The annulus surrounding the $\Gamma$-point in Figure~\ref{plot:graphene_bz}(c) may be related to the choice of energy range being somewhat lower than the $\pi-\sigma^*$ peak, as seen in Figure~\ref{plot:graphene_bz}(d).

In Figure~\ref{plot:phosphorene_bz}(a), (b), and (c) we show the weights of \Pn{}'s three main transitions over its orthorhombic BZ. Due to \Pn{}'s orthorhombic symmetry we expect its transitions to exhibit a four-fold symmetry over the BZ. Figure~\ref{plot:phosphorene_bz}(a) shows the  distribution of the oscillator strengths $f_{n n'\bk}^{\hbq_x}$ for an anisotropic excitation sensitive to $x$-polarized light at  $\omega_x \approx 1.9$~eV.  This energy is $\sim0.1$~eV above the peak (\emph{cf.}\ Figure~\ref{plot:phosphorene_bz}(d)) to better show the transition's anisotropy. Not only do we notice the expected centering behaviour around the $\Gamma$ high symmetry point, but also an anisotropy between the $k_x$ and $k_y$ directions, yielding an elliptical annulus in the BZ.

Figure~\ref{plot:phosphorene_bz}(b) and (c) show BZ plots for energies higher in the spectrum corresponding to $y$ dominant $\omega_{yx} \approx 5.1$~eV and a second $x$-polarized $\omega_{2x} \approx 9.5$~eV transition, respectively.  The weak yet scattered density of $f_{n n'\bk}^{\hbq_x}$ suggests contributions from many bands with clear $k_x$ and $k_y$ dependencies. Ellipse-like shapes also appear, and give some indication of which $k$-points in the BZ are most relevant for a given transition energy, information that is not clearly evident from the  band structure alone (\emph{cf.}\ Figure~\ref{plot:phosphorene_bz}(e)). 
\begin{figure}
	\includegraphics[width=\columnwidth]{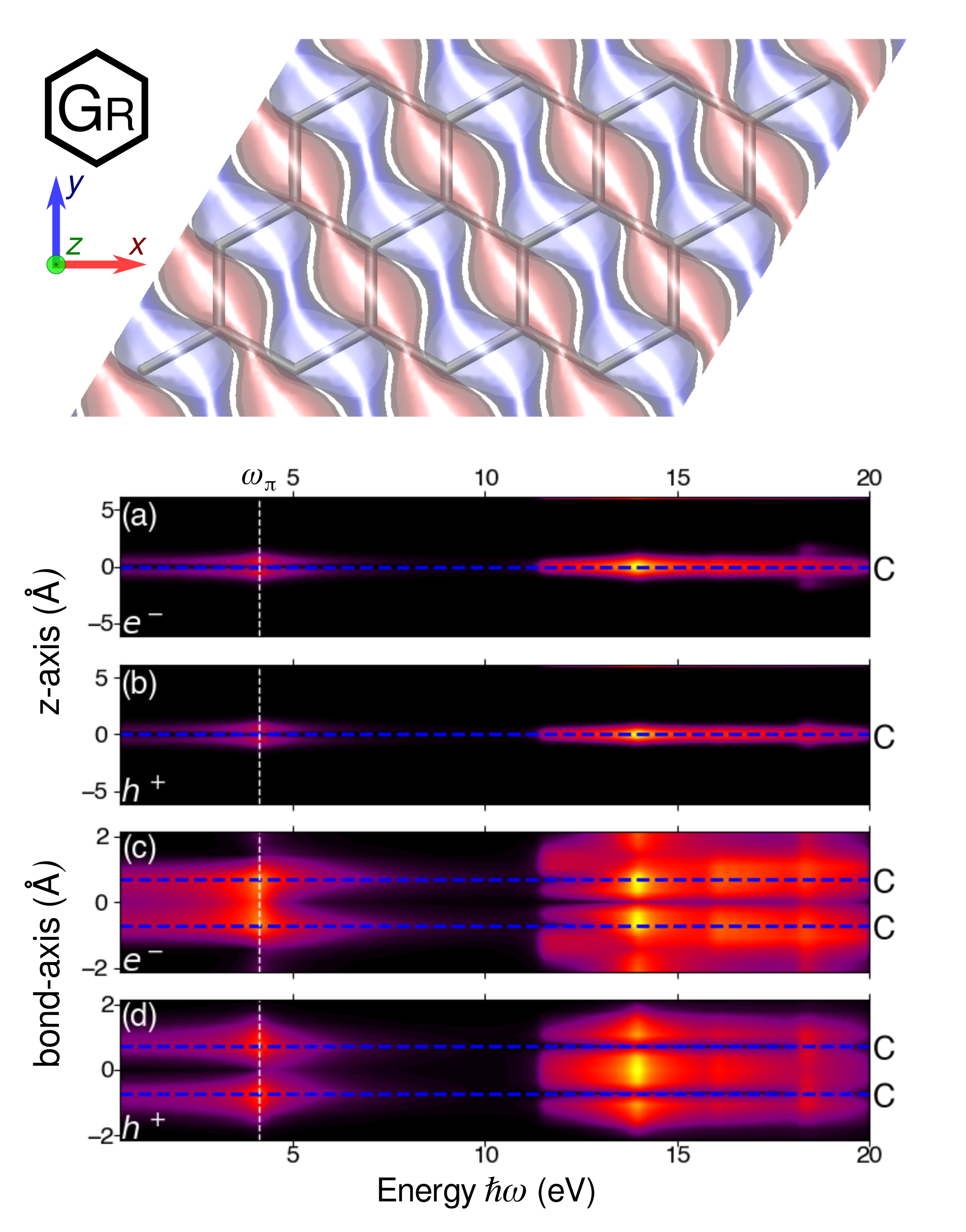}
	\caption{Graphene (\Gr) \LCAOTDDFTkomega{} calculated (a,c) electron ($e^-$) and (b,d) hole ($h^+$) densities, $\rho_e$ and $\rho_h$, and C atoms (blue dashed lines) projected onto the (a,b) $z$ or (c,d) bond axis versus energy $\hslash\omega$ in eV for in-plane polarized light, and positive (red) and negative (blue) isosurfaces of the $\pi - \pi^*$  exciton's density difference $\Delta\rho (\br, \hbq_x, \omega_\pi) =\rho_e (\br, \hbq_x, \omega_\pi) + \rho_h (\br, \hbq_x, \omega_\pi)$ at $\hslash\omega_\pi = 4.1$~eV (white dashed line) with axes shown as an inset.}\label{plot:graphene_real}
\end{figure}

\begin{figure*}
	\includegraphics[width=1.5\columnwidth]{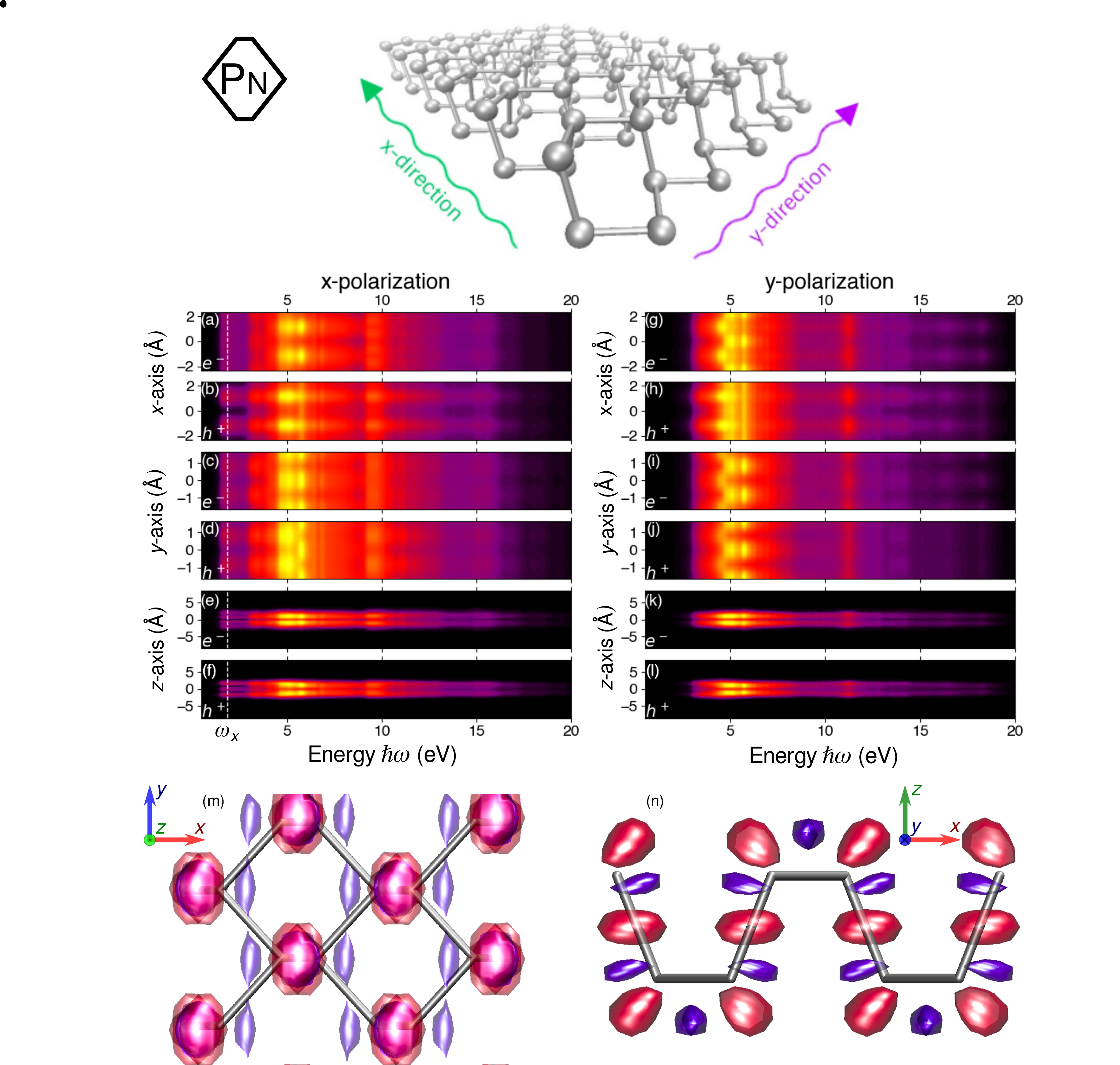}
	\caption{Phosphorene (\Pn) \LCAOTDDFTkomega{} calculated (a,c,e,g,i,k) electron ($e^-$) and (b,d,f,h,j,l) hole ($h^+$) densities, $\rho_e$ and $\rho_h$, projected onto the (a,b,g,h) $x$, (c,d,i,j) $y$, or (e,f,k,l) $z$ axis versus energy $\hslash\omega$ in eV for (a--f) $x$ and (g--l) $y$ polarized light, as depicted schematically in the upper inset, and (m,n) positive (red) and negative (blue) isosurfaces of the $x$-polarized exciton's density difference $\Delta\rho (\br, \hbq_x, \omega_x) = \rho_e (\br, \hbq_x, \omega_x) + \rho_h (\br, \hbq_x, \omega_x)$ at $\hslash\omega_x = 1.8$~eV (white dashed line) with axes shown as insets.}\label{plot:phosphorene_real}
\end{figure*}

As given in \eqref{eqn:twopoint}, the oscillator strengths $f_{n n'\bk}^{\hbq}$ derived in the \LCAOTDDFTkomega{} method can also be used to define a two-point excitonic density $\rhoex (\br_e,\br_h,\hbq, \omega)$ from which position-resolved hole and excited electron densities, $\rho_h (\br_h, \hbq, \omega)$ and $\rho_e (\br_e,\hbq, \omega)$, can be obtained from \eqref{eqn:rhoh} and \eqref{eqn:rhoe} by integration over the electron and hole positions, respectively. Figures~\ref{plot:graphene_real} and \ref{plot:phosphorene_real}, show the separate electron and hole densities and their difference. 

In Figure~\ref{plot:graphene_real} we show four real-space contours of either the electron or hole density projected onto either the $z$-axis, summing over the $x y$-plane, or along the C--C bond, summing over the plane normal to the C--C bond. From Figure~\ref{plot:graphene_real}(a) and (b) we can clearly see that both the electron and hole densities at $\hslash\omega_\pi  \approx 4.1$~eV have nodal planes in the \Gr{} plane, consistent with $\pi^*$ and $\pi$ orbitals, whereas the electron and hole densities at $\hslash\omega_\sigma \approx 14.1$~eV are localized within the \Gr{} plane, consistent with $\sigma^*$ and $\sigma$ orbitals.  The peak at $16.3$~eV is more difficult to analyze, although the lack of a nodal plane  at that energy suggests this transition should have some $\sigma$ character.

The $18$~eV peak of Figure~\ref{plot:lcao} in the \LCAOTDDFTkomega{} spectra has an excited electron with a significant spatial extent into the vacuum out of the \Gr{} plane (\emph{cf.}\ Figure~\ref{plot:graphene_real}(a)).  This suggests the excitation could be to a Rydberg state \cite{Echenique_1978}. However, this peak is absent in both the \PWTDDFTkomega{} and SZP \LCAOTDDFTkomega{} spectra (\emph{cf.}\ Figure~\ref{plot:lcao}).  Moreover, a proper description of diffuse Rydberg states, which sample the long-range area of the Coulomb potential tail, is not viable in standard TDDFT implementations \cite{Huser_2013}.  Altogether, this suggests this peak is an artifact of the basis set choice.  

Figure~\ref{plot:graphene_real}(c) and (d) yield spatially-resolved information about an excitation's electron and hole, respectively, along the C--C bonds within \Gr's unit cell. Again, $\pi^*$ and $\pi$ behavior is clearly seen at the $4.1$~eV peak, with the electron density shared between the two C atoms and the hole density centred on the atoms themselves. The opposite behavior is seen for the $14.1$~eV $\sigma - \sigma^*$ transition, with the spatial extent of the density stretching to the edges of the unit cell. 

The inset of Figure~\ref{plot:graphene_real} shows the electron and hole density difference $\Delta\rho (\br, \hbq_x, \omega_\pi)$ for the $\pi - \pi^*$ exciton at $\hslash\omega_\pi   \approx 4.1$~eV. We see a noticeable directional dependence of $\Delta\rho (\br, \hbq_x, \omega_\pi)$.  While the projections of the electron and hole densities must preserve \Gr's in-plane and out-of-plane symmetries, the density distributions will necessarily depend on the initial choice of polarization direction $\hat{\bq}$ relative to the material, although the calculation of $\Delta\rho (\br, \hbq_y, \omega_\pi)$ would yield a rotationally isomorphic density distribution. These striped isosurfaces are more reminiscent of the (10, 0) semiconducting SWCNT shown in Figure~\ref{fig:100SWCNT}, with the hole density along the zigzag direction, than the (10,10) metallic SWCNT shown in Figure~\ref{fig:1010SWCNT}.  Since \Gr{} is generally isotropic, we have equivalent electron and hole densities for any in-plane polarization direction. This is not the case for an anisotropic material such as \Pn.

Figure~\ref{plot:phosphorene_real} shows projections onto the $x$, $y$, and $z$ axes for $\hat{\bq}$ in both the $x$ and $y$ directions, as depicted schematically in the inset. From \Pn's $x$ and $y$ projections we see both the electron and hole extend throughout the unit cell.  This can be attributed to the dispersed locations of P atoms within the unit cell. We also see an anti-node/node in the $x$-polarization (\emph{cf.}\ Figure~\ref{plot:phosphorene_real}(a) and (b)) for the $x$-axis projection at $\hbar\omega_x \approx 1.8$~eV, reminiscent of the $\pi - \pi^*$ transition in graphene.  Otherwise, the two polarization directions look nearly identical for both electron and hole densities for in-plane projections. The $z$-axis projections show that the spatial extent of the electron and hole densities do not go far beyond \Pn's crystal structure, as we also saw for \Gr. We also see a nodal behaviour that persists in the $z$-projections.  This suggests the electron density is low near the centre of out-of-plane bonds. 

The electron-hole density difference $\Delta\rho (\rv, \hbq_x, \omega_x)$ is shown in Figure~\ref{plot:phosphorene_real}(m) and (n) for $x$-polarized light at $\hslash\omega_x \approx 1.8$~eV.  This is because for $y$-polarized light this exciton is noticeably dark (\emph{cf.}\ Figure~\ref{plot:phosphorene_real}(g--l)). It is interesting to note that for this polarization, the electron density appears to reveal stripes along the $y$-direction, while the hole density remains concentrated along the out-of-plane bonds. As a result, dipoles pointing between opposite charge densities would point in the $x$-direction, matching the calculated conductivity $\sigma(\omega)$ of Figure~\ref{plot:phosphorene_conductivity}.

\subsection{3D Anatase \& Rutile TiO$_{\text{2}}$}

Turning now to 3D bulk materials, we will assess the performance of the \LCAOTDDFTkomega{} method for describing the dielectric function $\varepsilon(\omega)$ of two prototypical photocatalytic materials: anatase (\ATiO) and rutile (\RTiO) titania.  These materials are excellent candidates for benchmarking the \LCAOTDDFTkomega{} method due to their technological relevance for photovoltaic and photocatalytic applications\cite{TiO2PhotocatalysisChemRev2014}, the plethora of literature describing their properties, and the difficulties in modelling their optical spectra from both a computational and theoretical point of view\cite{OurJACS}.  Specifically, the important role played by Ti's unoccupied $d$ levels in determining the band gap, absorbance, exciton binding, and electron-hole gemination in \ATiO{} and \RTiO{} make these materials a challenge to describe, often requiring quasiparticle calculations at the $G_0W_0$-BSE level.\cite{MiganiLong,mowbraytio2}.

\begin{figure}
	\includegraphics[width=\columnwidth]{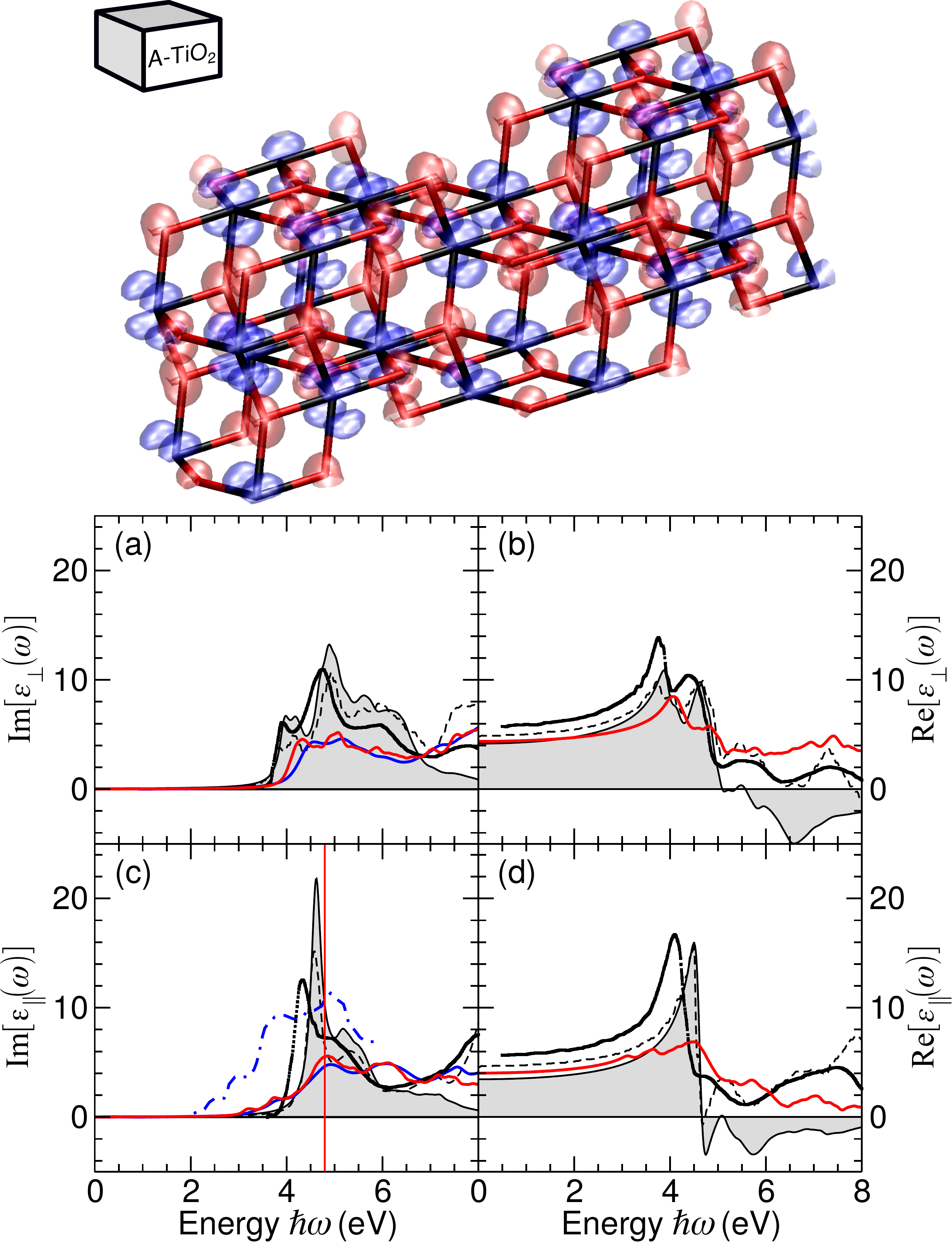}
	\caption{Anatase (\ATiO) (a,c) imaginary and (b,d) real parts of the dielectric function $\varepsilon(\omega)$ (a,b) perpendicular ($\perp$) and (c,d) parallel ($\|$) to the tetragonal $c$-axis versus energy $\hslash \omega$ in eV from reflectometry (black square)\cite{TiO2AnataseReflectivityDielectricConstantExp}, $G_0 W_0$-BSE (thin solid filled\cite{mowbraytio2} and dashed\cite{Landmann_2012} black lines), \PWTDDFTkomega{} (dash-dotted\cite{Chiodo_TiO2} and solid blue lines), and \LCAOTDDFTkomega{} (red solid lines), with positive (red) and negative (blue) isosurfaces of the bright exciton's density difference $\Delta\rho (\br, \hbq_x, \omega_{\textit{ex}}) = \rho_e (\br, \hbq_x, \omega_{\textit{ex}}) + \rho_h (\br, \hbq_x, \omega_{\textit{ex}})$ at (c) $\hslash\omega_{\textit{ex}} \approx 4.8$~eV (red vertical line) from \LCAOTDDFTkomega{}, shown as an inset, with O and Ti atoms colored red and grey, respectively.}\label{plot:anatase}
\end{figure}

\begin{figure}
	\includegraphics[width=\columnwidth]{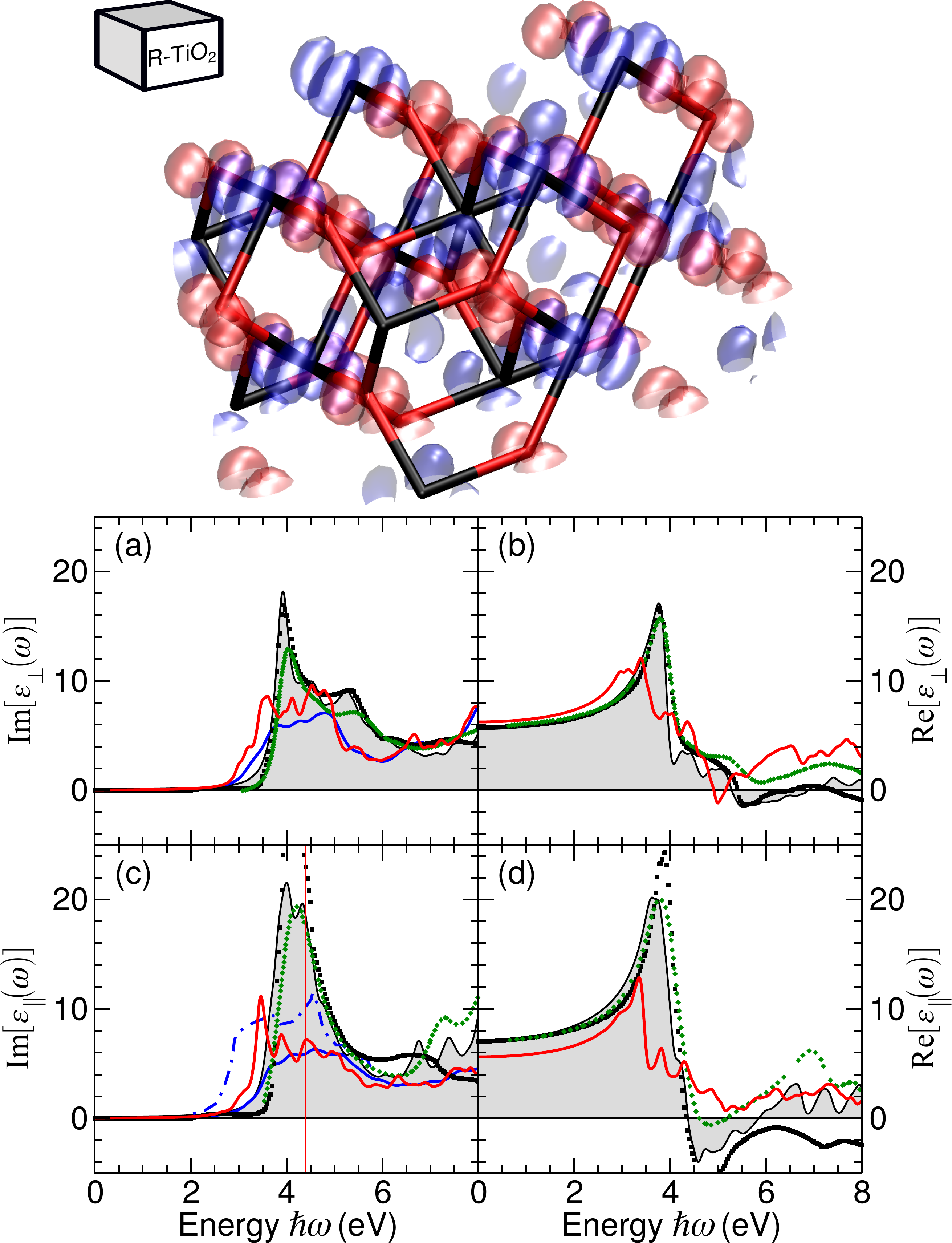}
	\caption{Rutile (\RTiO) (a,c) imaginary and (b,d) real parts of the dielectric function $\varepsilon(\omega)$ (a,b) perpendicular ($\perp$) and (c,d) parallel ($\|$) to the tetragonal $c$-axis versus energy $\hslash \omega$ in eV from reflectometry (black squares)\cite{PhysRev.137.A1467}, ellipsometry (green diamonds)\cite{tiwald_2000}, $G_0 W_0$-BSE (thin solid filled lines)\cite{MiganiLong}, \PWTDDFTkomega{} (dash-dotted\cite{Chiodo_TiO2} and solid blue lines), and \LCAOTDDFTkomega{} (red solid lines), with positive (red) and negative (blue) isosurfaces of the bright exciton's density difference $\Delta\rho (\br,\hbq_x, \omega_{\textit{ex}}) = \rho_e (\br, \hbq_x, \omega_{\textit{ex}}) + \rho_h (\br, \hbq_x, \omega_{\textit{ex}})$ at (c) $\hslash\omega_{\textit{ex}} \approx 4.4$~eV (red vertical line) from \LCAOTDDFTkomega{}, shown as an inset, with O and Ti atoms colored red and grey, respectively.}\label{plot:rutile}
\end{figure}

Figures \ref{plot:anatase} and \ref{plot:rutile} show the dielectric functions both parallel $\varepsilon_\|(\omega)$  and perpendicular $\varepsilon_{\perp}(\omega)$ to the tetragonal $c$-axis for \ATiO{} and \RTiO{}, respectively.
Both figures compare experimental reflectometry \cite{TiO2AnataseReflectivityDielectricConstantExp,PhysRev.137.A1467} or ellipsometry\cite{tiwald_2000} measurements with $G_0 W_0$-BSE, \PWTDDFTkomega{} and \LCAOTDDFTkomega{} calculations of the real and imaginary parts of the dielectric function $\varepsilon(\omega)$. $G_0W_0$-BSE and \PWTDDFTkomega{} calculations are included to determine the role played by many body effects, LCFs, and exciton binding in the description of \ATiO{} and \RTiO{}'s optoelectronic response functions. It is also important to see to what extent the more computationally efficient \LCAOTDDFTkomega{} method can describe the measured spectra. 

As can be seen for both \PWTDDFTkomega{} and \LCAOTDDFTkomega{} curves in Figures~\ref{plot:anatase} and \ref{plot:rutile}, due to the relatively flat $d$-orbital bands that compose the conduction band minima (CBM) for these polymorphs of TiO$_{2}$, the dielectric function is characteristically flat.

At the quasiparticle $G_0W_0$-BSE level, we find the excitonic binding is quite weak ($\sim0.1$~eV) in both \ATiO{}\cite{mowbraytio2} and \RTiO{}\cite{MiganiLong}, so that the positions of the first bright excitons are reasonably consistent with our \PWTDDFTkomega{} and \LCAOTDDFTkomega{} results, as shown in Table~\ref{tbl:TiO2}.
\begin{table}
  \caption{Energy of the first bright exciton $\hslash\omega_{\textit{ex}}$ in eV for anatase (\ATiO{}) and rutile (\RTiO).}\label{tbl:TiO2}
  \begin{ruledtabular}
    \begin{tabular}{lcc}
     & \multicolumn{2}{c}{$\hslash\omega_\textit{ex}$ (eV)}\\\cline{2-3}
		Method & \ATiO{} & \RTiO{}\\
		\cmidrule[0.2mm]{1-3}
		Reflectometry & $4.31$\footnote{Ref.~\citenum{TiO2AnataseReflectivityDielectricConstantExp}}  & $4.09$\footnote{Ref.~\citenum{PhysRev.137.A1467}}\\
		Ellipsometry & --- & $4.22$\footnote{Ref.~\citenum{tiwald_2000}}  \\
		$G_0 W_0$-BSE & $4.63$\footnote{Ref.~\citenum{mowbraytio2}} & $4.01$\footnote{Ref.~\citenum{MiganiLong}\nocite{MiganiLong}}\\
		$G_0 W_0$-BSE & $4.61$\footnote{Ref.~\citenum{Landmann_2012}}& $4.12$\footnotemark[6]\\
		\PWTDDFTkomega{} & $4.93$\footnote{Ref.~\citenum{Chiodo_TiO2}}  & $4.55$\footnotemark[7]\\
		\PWTDDFTkomega{} & $4.90$\footnote{This work.} & $4.56$\footnotemark[8]  \\
		\LCAOTDDFTkomega{} & $4.81$\footnotemark[8] & $3.48$\footnotemark[8]  \\
	\end{tabular}
  \end{ruledtabular}
\end{table}
Moreover, the \LCAOTDDFTkomega{} and \PWTDDFTkomega{} methods do qualitatively peak and trough with the experimental data (\emph{cf.}\ Figures~\ref{plot:anatase} and \ref{plot:rutile}).

The inclusion of exchange and correlation effects in the kernel $f_{\textit{xc}}$, beyond the Coulomb kernel $v$, allow quasiparticle $G_0W_0$-BSE calculations\cite{mowbraytio2,MiganiLong} to reproduce the spectral intensities semi-quantitatively.  For both \ATiO{} and \RTiO{} the dielectric functions obtained via \LCAOTDDFTkomega{} are generally more intense than their \PWTDDFTkomega{} equivalents. Finally, we expect the flat bands of TiO$_2$ will not exhibit a strong dependence on the choice of LCAO basis set or $k$-point sampling.

The electron-hole density difference $\Delta\rho (\br, \hbq_x, \omega_{\textit{ex}})$ for the bright excitonic peak $\hslash\omega_{\textit{ex}}$ in $\Imag[\varepsilon(\hbq_\|, \omega)]$ for \LCAOTDDFTkomega{} are marked and shown in Figures~\ref{plot:anatase} and \ref{plot:rutile}. We note, as in the case of graphene, that the symmetry of the in-plane response functions means that $\Delta\rho (\br, \hbq_y, \omega_{\textit{ex}})$ will be isomorphic to $\Delta\rho (\br, \hbq_x, \omega_{\textit{ex}})$. In both cases, the exciton electron is predominantly located on $d$-orbitals of the Ti atoms. For \ATiO{}, these orbitals all point in the same direction, whereas for \RTiO{} there are two orientations of the electron orbitals depending on the Ti atom's site. The hole, by contrast, is centred on the O atoms.  It is clearly seen from Figure~\ref{plot:rutile} that the hole has $2p$-character for the \RTiO, whereas for \ATiO{} (\emph{cf.}\ Figure~\ref{plot:anatase}) the hole has a mixture of $s$ and $p$-orbital character but is still centred on the O atoms. This is consistent with previous $G_0W_0$ results\cite{mowbraytio2} which found these excitations are predominantly from O 2$p$ occupied to Ti 3$d$ unoccupied orbitals.

\section{CONCLUSIONS}\label{Sect:Conclusions}

In this work, we have performed a thorough benchmarking of our \LCAOTDDFTkomega{} code\cite{code}, which uses a highly efficient linear combination of atomic orbitals (LCAOs)\cite{GPAWLCAO} to represent the KS wavefunctions, and performs TDDFT calculations in reciprocal space ($k$) and frequency ($\omega$) domains, while restricting calculations to the optical limit $\|\bq\| \to 0^+$. Specifically, we consider a large class of low-dimensional materials, namely C$_{60}$  (0D), metallic (10,10) and semiconducting (10,0) SWCNTs (1D), \Gr{} and \Pn{} (2D), and bulk \RTiO{} and \ATiO{} (3D). Our calculations agree qualitatively and semi-quantitatively with \PWTDDFTkomega{}, $G_0W_0$-BSE, and experimental measurements, while reducing the computational cost and improving stability relative to similar TDDFT methods. Working in the optical limit also provides direct access to the exciton's spatial distribution within the \LCAOTDDFTkomega{} method, and we demonstrate in this work how real space electron and hole density distributions, their difference, both in 3D and 2D projections, and the weights of the transitions in the BZ, provide a better understanding of energy-resolved spatial and reciprocal space excitation profiles. The significant reduction in computational cost when using the \LCAOTDDFTkomega{} method, combined with its accuracy, opens the door to future studies utilizing this method to perform computational screening studies of optoelectronic \cite{Optoelectronics2012,MolecularPlasmonicsACSPhotonics2019}, photovoltaic \cite{PolymerFullerenePhotovoltaic,FullereneOnlyPhotovoltaic}, and photocatalytic\cite{TiO2PhotocatalysisChemRev2014,OurJACS} systems \emph{in silico}.

\acknowledgments

This work was made possible by the facilities of the Shared Hierarchical Academic Research Computing Network (SHARCNET:\url{www.sharcnet.ca}) and Compute/Calcul Canada, employed the Imbabura cluster of Yachay Tech University, which was purchased under contract No.\ 2017-024 (SIE-UITEY-007-2017), and was supported by the QuantXLie Centre of Excellence, a project co-financed by the Croatian Government and European Union through the European Development Fund -- the Competitiveness and Cohesion Operational Program (Grant KK.01.1.1.01.0004).  K.L.\ and V.D.\ are grateful to Yachay Tech University and D.J.M. is grateful to Zoran Mi\v{s}kovi\'{c} and the University of Waterloo for their hospitality during various stages of this research. 

\appendix

\section{Derivative Discontinuity Correction}\label{section:ddc}

\begin{table}
  \caption{Derivative discontinuity correction $\Deltax$ in eV obtained from a linear combination of atomic orbitals (LCAO) or a plane wave (PW) representation of the KS wavefunctions for a fullerene (C$_{60}$), (10,0) SWCNT, phosphorene (\Pn), anatase (\ATiO), and rutile (\RTiO).}\label{tbl:Deltax}
  \begin{ruledtabular}
    \begin{tabular}{lcc}
       & \multicolumn{2}{c}{$\Deltax$ (eV)} \\\cline{2-3}
      Material & LCAO & PW \\\hline
      C$_{60}$ & 0.77 & ---\\
      (10,0) SWCNT & 0.79 & ---\\
      \Pn{} & 0.58 & 0.55\\
      \ATiO{} & 0.85 & 0.89 \\
      \RTiO{} & 0.67 & 0.72\\
  \end{tabular}
  \end{ruledtabular}
\end{table}

The derivative discontinuity correction to the exchange part of the \GLLBsc{}\cite{GLLB} functional $\Deltax$ provides an \emph{ab initio} first-order correction to the unoccupied KS eigenenergies using \eqref{eqn:Deltax}.  We have employed this correction for all semiconducting materials sudied herein.

As shown in Table~\ref{tbl:Deltax}, $\Deltax$ provides a substantial qualitative correction to the conduction bands of these materials, between 0.5 and 0.9~eV. Moreover, as seen from (\ref{eqn:q0limit}), $\Deltax$ not only shifts the peak positions in the calculated spectra, but also modifies the relative intensities of the peaks.  To account for this renormalization of the matrix elements in our \PWTDDFTkomega{} calculations, it has been necessary to rescale the calculated spectra by an \emph{ad hoc} $\left(\frac{\omega - \Deltax}{\omega}\right)^2$ factor, which yields semi-quantiative agreement with our \LCAOTDDFTkomega{} spectra (\emph{cf}. Figures~\ref{fig:100SWCNT}, \ref{plot:lcao}, \ref{plot:anatase}, and \ref{plot:rutile}).

\section{Irreducible Brillouin Zone (IBZ)}

From \eqref{eqn:chi} and \eqref{eqn:q0limit} it is clear that, by their definition, the optical response functions depend explicitly on the vector components of the wavenumber within the BZ. This means that, while it is often computationally convenient to work within the irreducible Brillouin zone (IBZ), such a naive application of the \LCAOTDDFTkomega{} method within the IBZ on a grid of $k$-points with non-trivial symmetries could yield incorrect response functions.

\begin{table}
  \caption{In-plane directional dependence of the dielectric function $\varepsilon(\omega)$ up to $20$~eV over irreducible (IBZ) and reducible (RBZ) Brillouin zones for graphene (\Gr), phosphorene (\Pn), anatase (\ATiO) and rutile (\RTiO). }
  \begin{ruledtabular}
  \begin{tabular}{@{\hspace{1em}}l@{\hspace{3em}}r@{.}l@{\hspace{2em}}r@{.}l@{\hspace{2em}}r@{.}l@{\hspace{2em}}r@{.}l@{\hspace{1em}}}\label{table:ibz}
          & \multicolumn{4}{c@{\hspace{2em}}}{$\Real\left[\int \left|\varepsilon_{xx} - \varepsilon_{yy}\right| d\omega\right]$}
          & \multicolumn{4}{c}{$\Imag\left[\int \left|\varepsilon_{xx} - \varepsilon_{yy}\right| d\omega\right]$}                   \\\cline{2-5}\cline{6-9}
          Material & \multicolumn{2}{c@{\hspace{2em}}}{IBZ} & \multicolumn{2}{c@{\hspace{2em}}}{RBZ} &\multicolumn{2}{c@{\hspace{2em}}}{IBZ} & \multicolumn{2}{c@{\hspace{2em}}}{RBZ}\\\hline

                \Gr & 18&505  & 3&037 & 13&817  & 1&824\\
		\Pn & 10&698 & 11&481 & 9&250 & 9&763  \\
		\ATiO & 13&643 & 0&766  & 12&126 & 0&695  \\
		\RTiO & 11&889 & 0&494 & 10&776 & 0&367 \\
	\end{tabular}
       \end{ruledtabular}
\end{table}

As a measure of the calculated in-plane anisotropy of the dielectric function, in Table \ref{table:ibz} we provide the real and imaginary parts of the absolute dielectric function difference between $x$ and $y$ components integrated up to 20~eV, i.e., $\int \left|\varepsilon_{xx} - \varepsilon_{yy}\right| d\omega$, for the 2D and 3D materials studied herein.  We neglect the 0D and 1D materials as they employed trivial $k$-meshes. With the exception of \Pn, all these materials should exhibit symmetry of the dielectric tensor in the $x y$-plane, i.e., be isotropic.

Overall, numerical and floating point errors and the incompleteness of the LCAO basis set account for the observed differences in the reducible Brillouin zone (RBZ) calculations for planar isotropic materials.  However, these results clearly show in their trends the need to account for the vectorial nature of the matrix elements in \eqref{eqn:q0limit}.   For this reason, we have averaged over the two directional components, $\varepsilon_{\|} = \frac{1}{2}\left(\varepsilon_{xx} + \varepsilon_{y y}\right)$, for the in-plane component of the dielectric function for the isotropic materials (\Gr, \ATiO, and \RTiO), which can be proven geometrically to yield the appropriate response functions.

\section{Spectral Convergence with \emph{k}-Point Spacing}

\begin{figure}
	\includegraphics[width=\columnwidth]{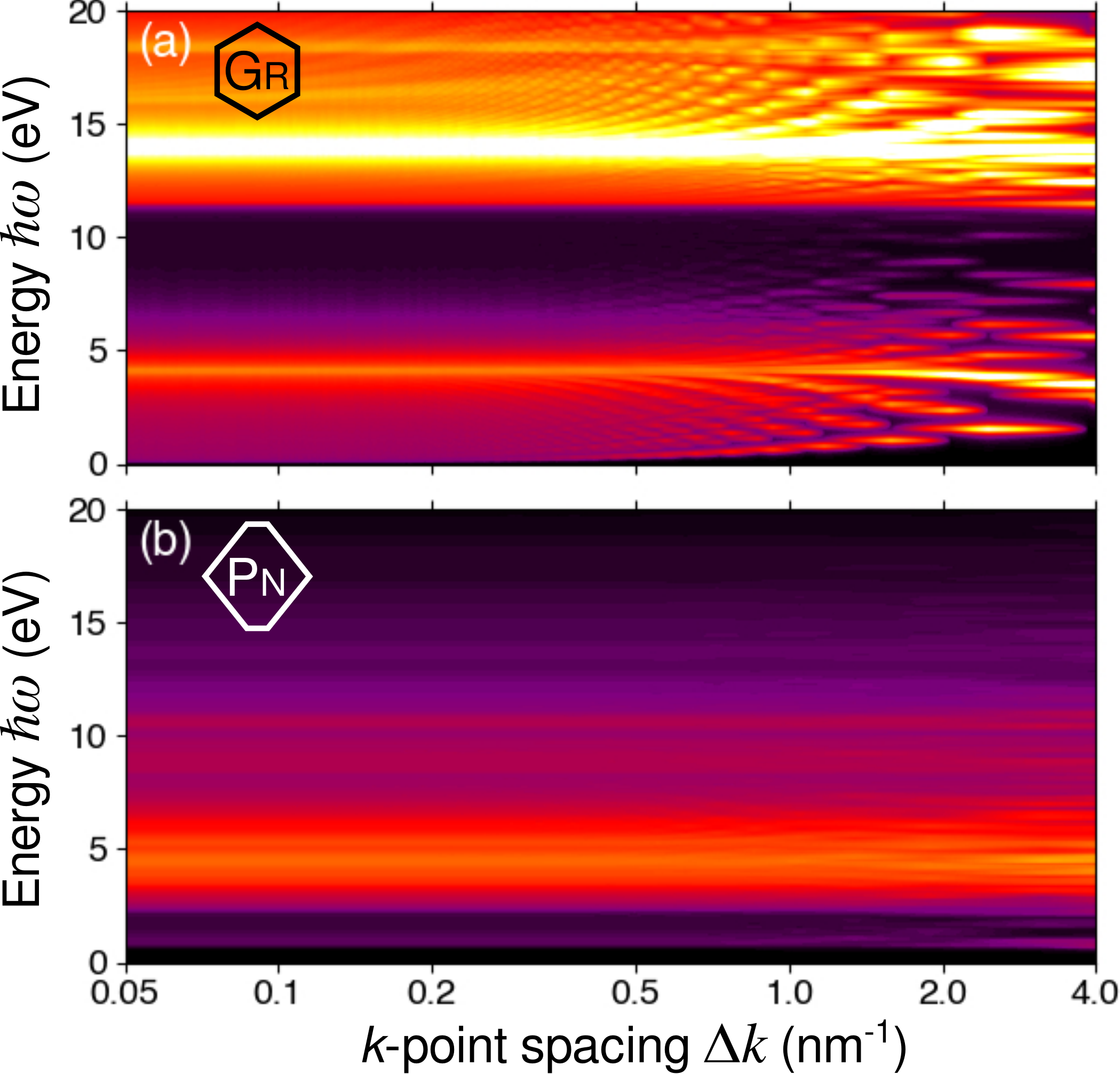}
	\caption{Convergence of real in-plane conductivity $\Real[\sigma(\hbq_\|, \omega)]$ spectra in eV with $k$-point spacing $\Delta k$ in (nm$^{-1}$) for (a) graphene (\Gr) and (b) phosphorene (\Pn) monolayers.}\label{plot:kmesh}
\end{figure}

In Figure~\ref{plot:kmesh} we plot the convergence of the real part of the in-plane conductivity $\sigma_\|(\omega)$ with $k$-point spacing $\Delta k$ for \Gr{} and \Pn{}.  These were chosen as they are the two materials considered herein with the greatest band dispersion.  For example, a $\Delta k_x = \Delta k_y \approx 0.05$~nm$^{-1}$ spacing for \Gr{} corresponds to a Monkhorst-Pack $541 \times 541 \times 1$ $k$-point mesh.

For \Gr, we find that most peaks in the real part of the conductivity are already converged for rather coarse grid spacings of $\Delta k \lesssim 0.5$~nm$^{-1}$, whereas the peak at $16.3$~eV only achieves full convergence with our \LCAOTDDFTkomega{} code\cite{code} for a very fine grid spacing of $\Delta k \lesssim 0.1$~nm$^{-1}$. This may be due to the fact that the $16.3$~eV transition takes place largely at the $\Gamma$ point, as seen in Figure~\ref{plot:graphene_bz}, which due to its high symmetry will be more sensitive to minute changes in the $k$-point mesh. For \Pn{} we find its spectra is basically converged for a coarse $\Delta k \lesssim 2$~nm$^{-1}$ grid spacing, with no substantial changes in the conductivity for denser $k$-point meshes.

\section{Implementation of Low-Dimensional Response Functions for Phosphorene}

We have used throughout optical response functions appropriate to the dimension of each material to better convey the importance of separating components of these functions depending on their parallel or perpendicular orientation. The use of appropriate response functions becomes apparent especially when dealing with layered materials. In this case, phenomenological models may treat an atomically thin material as either a thin 3D layer, where the 3D dielectric function should be used, or as a 2D boundary condition, where the 2D in-plane conductivity should be used.

\begin{figure}
	\includegraphics[width=\columnwidth]{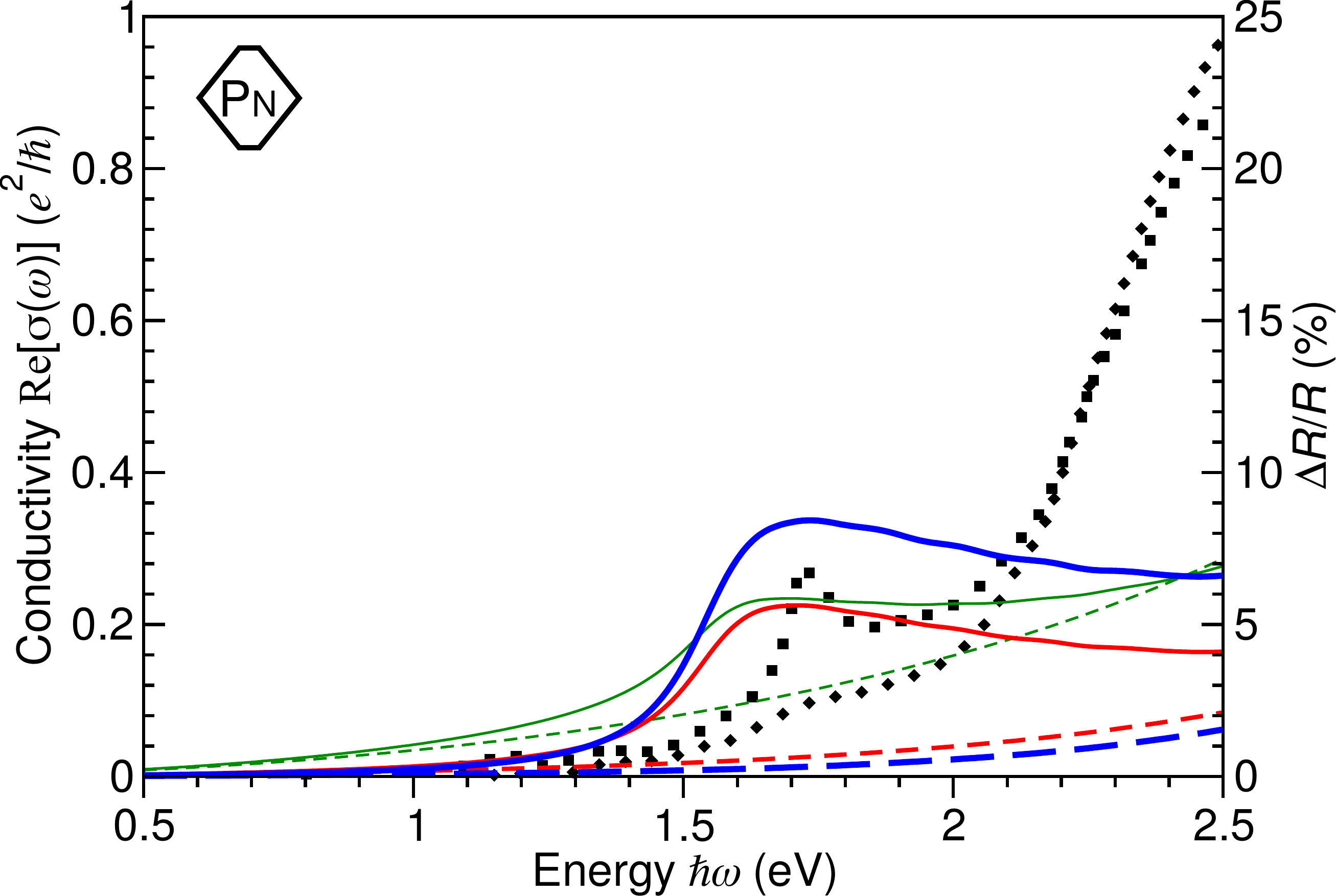}
	\caption{Phosphorene (\Pn) real part of the conductivity $\sigma(\omega)$ in $e^2/\hslash$ (blue thick lines) and reflection spectra $\Delta R/R$ in \% for 2D (red lines) and 3D (green thin lines) models and measurements (symbols)\cite{Li_2016}
          for light polarized in the $x$ (solid, squares)
          and $y$ (dashed, diamonds) directions.}\label{plot:reflection}
\end{figure}

In Figure \ref{plot:reflection}, we show experimental reflection spectra $\Delta R / R$ for \Pn{} sandwiched between a sapphire (\AlO) substrate and insulating hexagonal boron nitride (hBN) flakes of width $15$~nm for $x$ and $y$ polarized light\cite{Li_2016}. Also shown are two phenomenological models based on the results of our \LCAOTDDFTkomega{} calculations.  One is a 3D thin layer model for the air-hBN-\Pn-\AlO{} system based on the dielectric function $\varepsilon(\omega)$ computed via the \LCAOTDDFTkomega{} method\cite{Li_2016}. The other treats \Pn{} as a boundary condition with the 2D in-plane conductivity $\sigma_\|(\omega)$ computed via \eqref{eqn:conductivity} using Fresnel's equations through a transfer matrix formalism \cite{Lyon_2014}. 

The 2D approach yields a smaller tail compared to the 3D model, as well as a sharper peak, more in line with the experimental spectra\cite{Li_2016}. However, the 3D model better predicts the behaviour of the reflection spectra after the $1.8$~eV peak. It is important to note that the reflection ratio in the 3D model requires an empirical assumption about the thickness of the \Pn{} layer, to which it is directly proportional.  We employ a value of $0.5$~nm for this thickness, inferred based on the spacing between layers in multilayer phosphorene\cite{Li_2016}. The 2D model, by contrast, does not require any such assumption.  In this way it is a phenomenological model with fewer free parameters that agrees well with the experimental spectra\cite{Li_2016}. Although the use of the optical polarizability is not limited to the \LCAOTDDFTkomega{} method, it is used in our code\cite{code} to yield results theoretically more suited to lower-dimensional materials.

\bibliography{bibliography}

\end{document}